\documentclass[draft]{agujournal2019}

\usepackage{url} %this package should fix any errors with URLs in refs.
\usepackage{lineno}
\usepackage[finalnew]{trackchanges} %for better track changes. finalnew option will compile document with changes incorporated.
\usepackage{soul}

\draftfalse

\journalname{JGR: Space Physics}

\begin{document}

%% ------------------------------------------------------------------------ %%
%  Title
%
% (A title should be specific, informative, and brief. Use
% abbreviations only if they are defined in the abstract. Titles that
% start with general keywords then specific terms are optimized in
% searches)
%
%% ------------------------------------------------------------------------ %%

% Example: \title{This is a test title}

\title{Magnetic storms during the space age: Occurrence and relation to varying solar activity}

%% ------------------------------------------------------------------------ %%
%
%  AUTHORS AND AFFILIATIONS
%
%% ------------------------------------------------------------------------ %%

% Authors are individuals who have significantly contributed to the
% research and preparation of the article. Group authors are allowed, if
% each author in the group is separately identified in an appendix.)

% List authors by first name or initial followed by last name and
% separated by commas. Use \affil{} to number affiliations, and
% \thanks{} for author notes.
% Additional author notes should be indicated with \thanks{} (for
% example, for current addresses).

% Example: \authors{A. B. Author\affil{1}\thanks{Current address, Antartica}, B. C. Author\affil{2,3}, and D. E.
% Author\affil{3,4}\thanks{Also funded by Monsanto.}}

\authors{Kalevi Mursula\affil{1}, Timo Qvick\affil{1}, Lauri Holappa\affil{1}, Timo Asikainen\affil{1}}

% \affiliation{1}{First Affiliation}
% \affiliation{2}{Second Affiliation}
% \affiliation{3}{Third Affiliation}
% \affiliation{4}{Fourth Affiliation}

\affiliation{1}{Space Climate Group, Space Physics and Astronomy Research Unit, University of Oulu, Finland}
%(repeat as many times as is necessary)

%% Corresponding Author:
% Corresponding author mailing address and e-mail address:

% (include name and email addresses of the corresponding author.  More
% than one corresponding author is allowed in this LaTeX file and for
% publication; but only one corresponding author is allowed in our
% editorial system.)

% Example: \correspondingauthor{First and Last Name}{email@address.edu}

\correspondingauthor{Kalevi Mursula}{kalevi.mursula@oulu.fi}

%% Keypoints, final entry on title page.

%  List up to three key points (at least one is required)
%  Key Points summarize the main points and conclusions of the article
%  Each must be 140 characters or fewer with no special characters or punctuation and must be complete sentences

% Example:
% \begin{keypoints}
% \item	List up to three key points (at least one is required)
% \item	Key Points summarize the main points and conclusions of the article
% \item	Each must be 140 characters or fewer with no special characters or punctuation and must be complete sentences
% \end{keypoints}

\begin{keypoints}

\item We explain the occurrence of magnetic storms in space age by their relation to the varying solar activity and solar magnetic structure.
\item The occurrence of large, moderate and weak HSS/CIR storms follows the decrease of the HCS tilt in the declining phase of solar cycle.
\item Maxima of HSS/CIR storms have shifted from late declining phase in cycles 20-22 to earlier times in cycles 23-24 due to recent HCS widening.

% OTHER POSSIBLE 
% \item We use two indices to study magnetic storms during the space age, 1957--2021, dividing them according to intensity and solar wind source.
% \item First account of magnetic storms during the space age, 1957--2021, with storms divided according to intensity and solar wind source.
%\item While CME storms follow sunspots over the whole space age, HSS/CIR storms depend on the width and tilt of the HCS, maximizing in cycle 22.
% \item CME storms of different intensity follow sunspots at their specific relation, HSS/CIR storms depend on the width and tilt of the HCS, maximizing in cycle 22.
% \item Sunspots correlate with CME storms over the whole space age of different intensity follow sunspots at their specific relation, HSS/CIR storms depend on the width and tilt of the HCS, maximizing in cycle 22.
% \item HSS/CIR storm maxima move from the late declining phase in the early cycles 20--22 to the early declining phase in the recent weak cycles 23--24.

\end{keypoints}

%% ------------------------------------------------------------------------ %%
%
%  ABSTRACT and PLAIN LANGUAGE SUMMARY
%
% A good Abstract will begin with a short description of the problem
% being addressed, briefly describe the new data or analyses, then
% briefly states the main conclusion(s) and how they are supported and
% uncertainties.

% The Plain Language Summary should be written for a broad audience,
% including journalists and the science-interested public, that will not have 
% a background in your field.
%
% A Plain Language Summary is required in GRL, JGR: Planets, JGR: Biogeosciences,
% JGR: Oceans, G-Cubed, Reviews of Geophysics, and JAMES.
% see http://sharingscience.agu.org/creating-plain-language-summary/)
%
%% ------------------------------------------------------------------------ %%

%% \begin{abstract} starts the second page

\begin{abstract}

We study the occurrence of magnetic storms in space age (1957--2021) using Dst and Dxt indices.
We find 2526/2743 magnetic storms in the Dxt/Dst index, out of which 45\% are weak, 40\% moderate, 12\% intense and 3\% major storms.
Occurrence of storms in space age follows the slow decrease of sunspot activity and the related change in solar magnetic structure.
We quantify the sunspot - CME storm relation in the five cycles of space age.
We explain how the varying solar activity changes the structure of the heliospheric current sheet (HCS), and how this affects the HSS/CIR storms.
Space age started with a record number of storms in 1957--1960, with roughly one storm per week.
Solar polar fields attained their maximum in cycle 22, which led to an exceptionally thin HCS, and a space age record of large HSS/CIR storms in 1990s. 
In the minimum of cycle 23, for the only time in space age, CME storm occurrence reduced below that predicted by sunspots.
Weak sunspot activity since cycle 23 has weakened solar polar fields and widened the HCS, which has decreased the occurrence of large and moderate HSS/CIR storms.
Because of a wide HCS, the Earth has spent 50\% of its time in slow solar wind since cycle 23.
The wide HCS has also made large and moderate HSS/CIR storms occur in the early declining phase in recent cycles, while in the more active cycles 20--22 they occurred in the late declining phase.

\end{abstract}

\section{Introduction}
\label{Sec: Sec1 Introduction}

Magnetic storms \cite<for a review see, e.g.,>[]{Dessler_Parker_1959, Gonzalez_1994,Daglis_1999, Daglis_2003, Daglis_SSR_2006} are the largest disturbances of the near-Earth space, driven by enhanced interaction between the solar wind (SW), including the heliospheric magnetic field (HMF; also called the interplanetary magnetic field, IMF), and the Earth's magnetosphere.
Certain structures in the solar wind and the HMF contain favourable conditions for enhanced interaction that last typically from one to several days.
The two most important structures leading to storms are the interplanetary manifestations of coronal mass ejections (CME) \cite{Gosling_1993, Kilpua_LRSP_2017} and the corotating interaction regions (CIR) related to high-speed solar wind streams (HSS) \cite{Krieger_etal_73, Richardson_LRSP_2018}. 

CMEs arise from the eruption of large coronal flux tubes and lead to increased solar wind density and pressure \cite{Webb_LRSP_2012}.
They often have the internal structure of a magnetic cloud which includes a systematic rotation of magnetic field lines and a long period of southward directed magnetic field \cite{Burlaga_1981, Klein_1982}.
This leads to enhanced reconnection at the dayside magnetopause, producing a magnetic storm.
Coronal flux tubes that precede CMEs arise from the solar convection layer and have a close connection with sunspots.
Accordingly, the occurrence  of CMEs varies closely in phase with sunspot activity, reflecting the appearance of new magnetic flux on the solar surface \cite{Webb_and_Howard_1994, Gopalswamy_2004, Cremades_2007, Robbrecht_ASR_2006, Webb_LRSP_2012}.

High-speed solar wind streams emanate from large solar coronal holes \cite{Krieger_etal_73, Richardson_LRSP_2018}, where the magnetic field is typically unipolar and experiences less of super-radial expansion than in the neighborhood of closed field regions \cite{Wang_ApJ_1990}. 
Accordingly, solar wind flow can more freely escape from large coronal holes and is accelerated to higher speeds than elsewhere.
On the other hand, the wind originating at the streamer belt and neighboring regions is slower but more dense. 
Since the streamer belt also carries the heliospheric current sheet (HCS) and its center, the magnetic neutral line (NL), the slow wind is commonly equated with the HCS.
When the HCS is inclined with respect to the solar equator, the slow and fast solar wind regions can be at the same heliographic latitude and be emitted successively in the same direction.
When the fast stream catches the slow wind ahead, an interaction region called the corotating interaction region, also called the stream interaction region (SIR), is formed \cite{Balogh_1999, Gosling_1999}. 
Due to compressed plasma density and increased magnetic field intensity, as well as to the following fast solar wind, the HSS/CIR structure is very effective in producing weak and moderate storms \cite{Tsurutani_2006}.

Solar magnetic fields experience a dramatic structural change over the sunspot cycle.
During solar minima the global solar magnetic field is mainly dipolar, with a few active regions around the solar equator and large coronal holes of unipolar field with opposite polarities around the two solar poles \cite{Petrie_LRSP_2015}.
As activity increases in the ascending phase of the cycle, surges of magnetic flux with polarity opposite to the prevailing polar field are transported to each pole \cite{Wang_etal_1989, Virtanen_Mursula_ApJ_2014}.
These surges reduce the size of the old-polarity coronal holes and, eventually, reverse the polarity of the polar field around solar maxima.
Subsequent surges in the declining phase increase the new-polarity field at the poles and extend the area of polar coronal holes.
Moreover, surges can form contiguous unipolar regions from the low-latitude origin of surges to the solar pole.
This leads to the formation of longitudinally asymmetric extensions of polar coronal holes, so called elephant trunks, that can cover a wide range of latitudes and emit fast solar wind even at low latitudes \cite{Harvey_Sheeley_1979}.
Such extensions of polar coronal holes to low latitudes are an important source of high-speed streams reaching the Earth.
Smaller-scale coronal holes can be formed between active regions at any latitude and at any time of the cycle, especially at solar maxima.
Related HSS streams can affect (e.g., accelerate) CMEs bursting from neighboring active regions \cite{Crooker_Cliver_1994, Gosling_1996}.
Later in the declining phase, as the emergence of magnetic flux and related surge production subside, the global field slowly returns back to its minimum-time dipolar structure with a polarity structure opposite to the previous minimum.
The lower boundary of polar coronal holes becomes more symmetric, which reduces the tilt of the HCS and the occurrence of HSSs at the Earth.

In addition to this solar cycle evolution of solar magnetic fields, there are also longer-term changes, e.g., in the overall sunspot activity, leading to the varying height of sunspot cycles.
It is known that sunspot activity in the 20th century reached a record level at least for a couple of thousand years \cite{Solanki_2004}, culminating at the maximum of solar cycle 19 in 1957.
This highly active time of the mid-20th century is commonly called the Modern Grand Maximum (MGM).
After solar cycle 19 (to be called SC19), solar activity remained at a fairly high level for 3--4 cycles whereafter, since the maximum of SC23, solar activity has considerably subsided.
This is evidenced by an exceptionally long and deep minimum between SC23 and SC24, and a considerably low cycle 24.  
% One can call this 40-year time interval from 1957 to about 2000, which is marked by reducing solar activity, the declining phase of the GMM.
We note that, curiously, the start of space age, as marked by the flight of the first satellite in 1957, coincides with the maximum of the MGM. 
Thus, the space age, at least until the recent times, is characterized and affected by slowly, but unsteadily reducing solar activity.

The varying level of sunspot activity during the space age directly affects, e.g., the occurrence of CMEs and magnetic storms produced by CMEs.
In addition, there is also a related long-term change in the solar magnetic field structure, which affects coronal holes and, thereby, the occurrence of HSS/CIRs and magnetic storms produced by them.
In fact, while in the earlier, active cycles there were only small and short-lived coronal holes at low latitudes \cite{Fujiki_2016}, the declining phase of cycle 23 was characterized by rather large, persistent low-latitude coronal holes \cite{Gibson_2009,Hamada_SP_2021}, which led to a space age record activity of HSSs and geomagnetic activity in 2003 \cite{Mursula_2015}.
% The appearance of such low-latitude coronal holes is most likely related to the weakening of solar polar fields \cite{Smith_Balogh_2008,Zhou_2009}, as the result of reduced emergence of new flux on the solar surface.
Another long-term change is the recent weakening of solar polar fields \cite{Smith_Balogh_2008,Zhou_2009, Wang_2009} as the result of reduced emergence of new flux on the solar surface.
These changes have important consequences to the long-term occurrence of magnetic storms.
% Low-latitude coronal holes led to an all-time activity of HSSs and geomagnetic activity \cite{Mursula_2015} and it is likely that it has consequences also to geomagnetic storms.

The enhanced solar wind-magnetosphere interaction during CMEs and HSS/CIRs accelerates particles and drives several current systems that lead to magnetic disturbances in different parts of the Earth's surface. 
However, rather than measuring the overall disturbance level like, e.g., the Kp index of geomagnetic activity, magnetic storms are defined and quantified in terms of one current system only, the ring current \cite{Daglis_1999,Daglis_SSR_2006}.
The storm-time ring current consists mainly of (positively charged) hydrogen, helium and oxygen ions of some 10--300 keV energy, drifting around the Earth at the distance of about 3--7 Earth radii \cite{Daglis_1999}.
% While some other current systems are enhanced even by smaller intensifications of magnetospheric driving, the amount of ions forming the ring current is significantly enhanced only during the most persistent driving conditions.
% The storm-time ring current consists of (positively charged) hydrogen, helium and oxygen ions \cite{Daglis_1999} that drift around the Earth in the clockwise sense when viewed from the north pole.
Drifting westward around the Earth, they produce a negative deflection in the horizontal magnetic field on the ground, which is a direct measure of the energy content of the ring current \cite{Dessler_Parker_1959,Sckopke_1966}.

This deflection has been measured from 1957 onward by four ground-based magnetometer stations located at low latitudes, roughly equidistantly in longitude.
A dedicated recipe was developed \cite{Sugiura_1964,Sugiura_1969,Sugiura_1991,WDC_C2_Dst_2022} in order to remove the secular, seasonal and daily quiet-time variations from the locally observed magnetic field, and to quantify the ring current in terms of an index called the Dst index. 
Year 1957 was the International Geophysical Year (IGY), when several international programs on solar-terrestrial research were started, developing both ground-based and flying instrumentation.
% The Dst index is available since 1957, the International Geophysical Year (IGY), when several international programs on solar-terrestrial research were started, developing both ground-based and flying instrumentation.
% This development also led to the start of space era and the monitoring of space by satellites since the early 1960s.
Thus, it is not surprising that the Dst index is available since the start of space age.
However, as noted above, it is curious, but also quite appropriate, that the Sun reached its all-time maximum activity during the International Geophysical Year.
% Due to this coincidence magnetic storms can be monitored during the whole declining phase of the GMM by the Dst index.
% Note also that, because the IGY-inspired space era started fairly soon after the IGY, we have satellite-based observations of the solar wind and the HMF for most of the time of the Dst index.
Since continuous monitoring of space by satellites started soon after the IGY, we also have satellite observations of the solar wind and the HMF available for most of the time of the Dst index.
This allows us to study the solar wind drivers of magnetic storms almost over the whole space age.
%We will use these observations here extensively by assigning geomagnetic storms to their main drivers.

During the several years of storm studies some errors and inconsistencies have been found in the Dst index \cite{Karinen_2002,Mursula_etal_ASTRA_2008,Mursula_etal_JASTP_2011}.
Therefore \citeA{Karinen_Mursula_2005} recalculated the Dst index using the original recipe but correcting the noticed problems.
This revised Dst index is called the Dxt index.
We will use here both the Dst and the Dxt index in order to study magnetic storms during the space age in 1957--2021. 
Even though there are considerable differences between the two ring-current indices leading, e.g., to somewhat different numbers of magnetic storms, we find that the storm occurrences and their implications about the long-term change of the Sun, remain the same.

This paper is organized as follows. 
In Section~\ref{Sec: Sec2 Indices}, we present the Dst and Dxt indices and discuss their differences.
Section~\ref{Sec: Sec3 SW types} describes the solar wind classification into the three main structures (flow types).
Section~\ref{Sec: Sec4 Storm identification} explains how storms are derived from the two indices.
Section~\ref{Sec: Sec5 Storm numbers} gives the total numbers and mean intensities of storms of different intensity for the whole space age, and Section \ref{Sec: Sec6 Yearly numbers} presents their yearly numbers.
We assign the storms into their solar wind drivers in Section \ref{Sec: Sec7 Storm drivers} and present their yearly numbers in Section \ref{Sec: Sec8 Storm drivers Yearly}.
In Section~\ref{Sec: Sec9 CME storms} we discuss large and moderate CME storms and their relation to sunspots, separately during the five solar cycles.
Section~\ref{Sec: Sec10 HSS/CIR storms} presents the connection between HSS/CIR storms and the structure of the heliospheric current sheet, and discusses the implications of our results to the long-term evolution of the Sun.
Finally, in Section~\ref{Sec: Sec11 Conclusions} we discuss the obtained results and give our conclusions.

\section{Dxt and Dst indices}
\label{Sec: Sec2 Indices}

As noted above, in our quest for long-term homogeneity, we have found earlier that the Dst index depicts some errors and inconsistencies.
Correcting these problems eventually led to the recalculation of the index and to the development of the corrected Dst index, the Dxt index \cite{Karinen_Mursula_2005}. 
The first error found was an erroneous diurnal variation of the Dst index.
While the Dst index includes a rather small diurnal UT variation, 
% \citep{Mayaud-1978, Saroso-1993, Siscoe-1996, Cliver-2000}
% mainly due to the longitudinally inhomogeneous and latitudinally asymmetric distribution of the Dst stations \citep{Takalo-2001A}, 
it is artificially enlarged in 1971 \cite{Takalo_Mursula_2001,Takalo_ESA_UT_2002}.
We suggested \cite{Karinen_2002,Karinen_Mursula_2005} that this erroneous UT variation in 1971 is most likely due to an erroneous treatment of the diurnal variation of SJG station data.
Unfortunately, the WDC-C2 at Kyoto does not provide the individual disturbances of the four Dst stations, i.e., the local Dst indices, which would clarify this problem.
Therefore the cause of this error remains without final clarification.
Nonetheless, the Dst index is erroneous in 1971 and the Dxt index corrects this error.

The Dst index is some 2\,nT more negative, on an average, than the Dxt index \cite{Karinen_Mursula_2005}.
However, there are four consecutive years in 1963--1966 when the Dst index is considerably above (less negative than) the Dxt index \cite<see Figure 5 in>[]{Karinen_Mursula_2005}.
% The absolute difference between the Dst and Dxt indices is, on an average, about 2--3 nT.
% However, there are four consecutive years in 1963--1966 when this difference is enhanced three times larger \cite<see Figure 5 in >[]{Karinen_Mursula_2005}.
% In these years the Dst index is, by the same amount, larger (less negative) than the Dxt index.
% Year 1965 is unique in being the only year for which the annual Dst index is positive, far above any other year.
% However, annual means of the Dxt index remain negative in all years, even in 1965.
Moreover, it was found that the year 1965 is unique in that it is the only year for which the annual Dst index is positive, far above any other year.
In comparison, annual means of the Dxt index remain negative in all years, even in 1965.
Again, at the lack of local Dst indices, the cause of the exceptionally high Dst index in 1963--1966 remains without final explanation, but the great consistency of global and local Dxt indices suggests that the Dst index is slightly flawed in these years.

% The documentation of the derivation of the Dst index is partly deficient and even slightly ambiguous.
% Therefore the Dst index is not fully reproducible as such, even though one tries to follow the recipe as closely as possible.
% One ambiguity relates to the latitudinal normalization of the Dst index.
The latitudinal normalization of the Dst index is questionable.
Originally the mean of local disturbances was normalized by the cosine of the mean of geomagnetic latitudes \cite{Sugiura_1964}.
However, later the mean of local disturbances was normalized by the mean of the cosines of geomagnetic latitudes \cite{Sugiura_1991}.
Alas, since the Dst stations are at different latitudes, each local disturbance should be normalized by the cosine of the respective geomagnetic latitude in order to find the local disturbance of the same equatorial (horizontal) electric current.
Only then the different stations measure the same ring current intensity, and their normalized disturbances (local Dst indices) can be averaged to find the global mean of the ring current \cite{Hakkinen_JGR_2003,Karinen_Mursula_2005,Mursula_etal_ASTRA_2008}.
Note that this physically correct way of normalization has never been adopted in the Dst index recipe, nor can the Dst index be properly normalized because of the lack of local disturbances.
Note also that, although later studies have shown that other magnetospheric current systems \cite{Burton_1975, OBrien_2000, Asikainen_2010} and induced earth-currents also contribute to the Dst index, their contribution is not subtracted from the Dst index when identifying storms or estimating their intensity.

Here we mostly follow the Dxt reconstruction presented in Karinen and Mursula (2005).
However, meanwhile, we have adopted a few new practices that slightly modify the original Dst recipe and further improve the Dxt index.
In order to reduce the effect of seasonal variation and solar activity upon the baseline, we now calculate the secular variation using local midnight values (23 and 00 in local time) of the international quiet days. 
We also use a more sophisticated smoothing of the annual values by a 5-point quadratic Savitzky-Golay filter \cite{Savitzky_Golay_1964}, instead of using the second-order polynomial of the original recipe. 
We now calculate the hourly values of the baseline by interpolating the smoothed annual values to hourly resolution by a piecewise cubic Hermite interpolating polynomial (PCHIP). 
We also smooth the daily quiet-time curves using a 60-day Gaussian window, which removes the steps between successive monthly values that exist in the Dst index and in the earlier version of the Dxt index.

\begin{figure}[h!]
	\centering
	\includegraphics[width=1\linewidth]{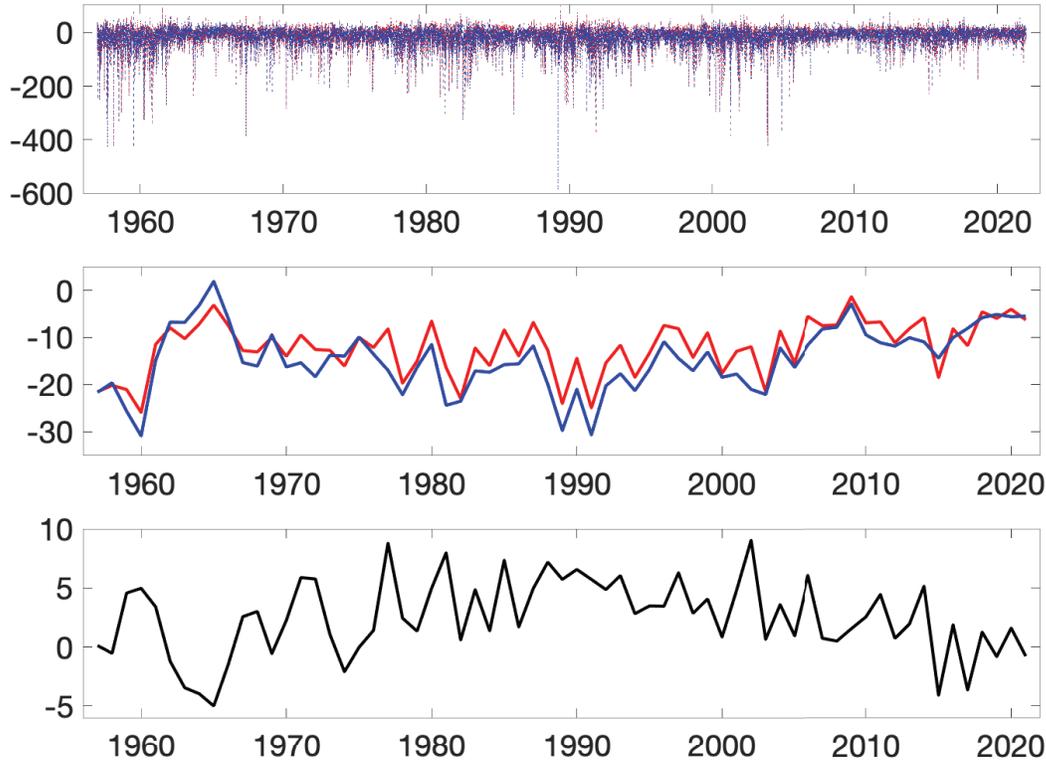}
	\caption{Hourly (top panel) and yearly (middle panel) values of the Dxt (red) and Dst (blue) indices, and the Dxt-Dst difference (bottom panel) of yearly values.}
	\label{fig:Dxt_Dst_hourly_yearly}
\end{figure}

The top and middle panels of Figure \ref{fig:Dxt_Dst_hourly_yearly} show the hourly and annual values of the Dxt and Dst indices, and the bottom panel depicts the Dxt-Dst difference of annual indices. 
One can see that the two indices differ even at annual resolution where their mean difference (mean absolute difference) is about 2.5\,nT (3.3\,nT, respectively). 
Note that the Dxt-Dst difference depicts a systematic long-term variation, with larger positive differences (relatively more disturbed Dst) during the active times from 1970s to early 2010s, and larger negative differences (relatively more disturbed Dxt) during the less active times (mainly in 1960s).
%and during the last 15 years. 
This inconsistent long-term difference between the two indices is mainly due to the inconsistent quiet-time level of the Dst index, which follows the varying level of solar activity, being too low in the weak decade of 1960s and too high in the subsequent, more active decades. 
While a more complete analysis of the differences between the two indices will be presented elsewhere, here we will only discuss the practical consequences of these differences to the number and strength of magnetic storms.
% While the quiet-time level of the Dxt index has a roughly constant long-term level, the quiet-time level of the Dst index follows closely the long-term evolution of the Dxt-Dst difference.
In particular, because the Dst index is, on an average, 2.5\,nT lower than the Dxt index, it produces a somewhat larger number of magnetic storms than the Dxt index.
However, as we will show in this paper, despite these differences, our results on storm occurrences and their implications about the long-term change of the Sun, remain the same for the two indices.

% FIGURE1

We would also like to note that there are currently a few slightly different versions of the Dst index in the different servers and databases. 
E.g., the Dst index in the ISGI data server and in the OMNI2 database of the NSSDC differ from the Kyoto WDC Dst index and from each other during long time intervals in the last few years, at least by the time of this writing (April 2022).
This is most likely due to different (provisional or final) versions of the Dst index of WDC being implemented in the other databases at different times.
(We have notified the respective institutions of these differences. 
A detailed list of these differences can be obtained from the authors of this paper at request.)
Finally, we note that there are a number of more recent developments aiming to an improved estimate of the ring current, either using a novel reconstruction method \cite{Love_2009}, improved temporal accuracy \cite{Gannon_2011} or increased spatial accuracy by an extended station network \cite{Newell_2012}.
While all of these developments are motivated and have their own applications, we will use here only the Dst/Dxt indices since they are the only indices to cover the whole space age.

%\newpage
\section{Solar wind flow types}
\label{Sec: Sec3 SW types}
 
\citeA{Richardson_JGR_2000,Richardson_etal_2002a} and \citeA{Richardson_GA_2012} developed an hourly list of solar wind flow types using the measured values of the near-Earth solar wind parameters and some auxiliary data, like the sudden storm commencements, magnetospheric energetic particles, and cosmic rays. 
They classified the solar wind into three different main flow types: 
\begin{enumerate}
\item CME-related flows that consist of interplanetary CMEs, including their possible upstream shocks and sheath regions											
\item high-speed streams and the related corotating interaction regions											
\item slow solar wind that can be related with the streamer belt and the HCS.											
\end{enumerate}

\begin{figure}[h!]
	\centering
	\includegraphics[width=1\linewidth]{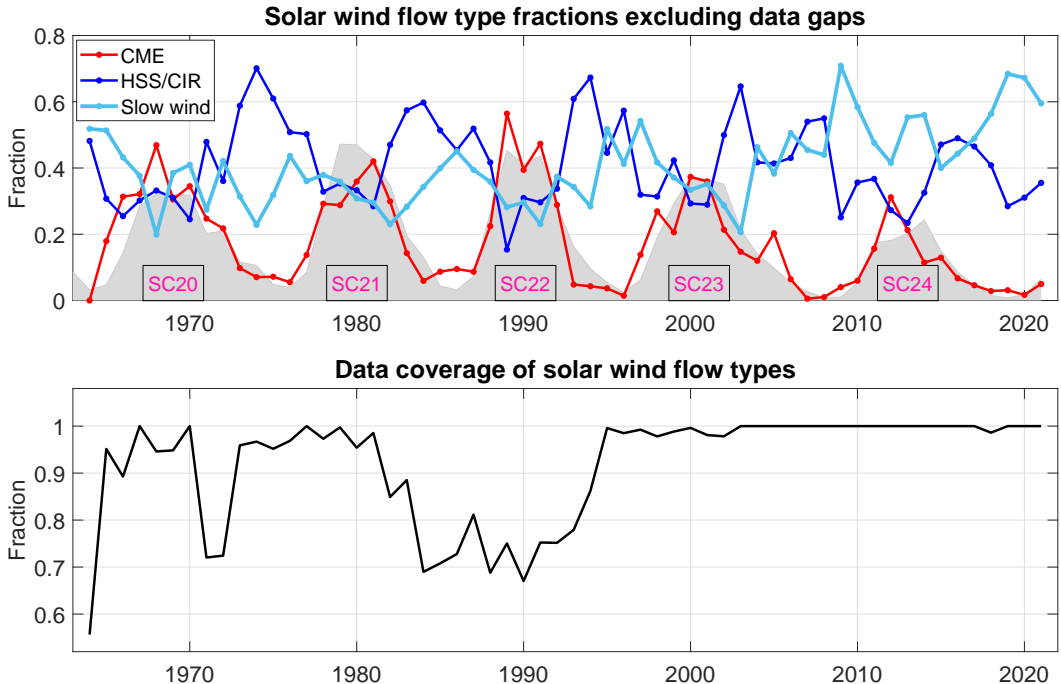}
	\caption{Top: Annual fractions of CME (red), HSS/CIR (blue) and slow solar wind (cyan) solar wind flows in 1964--2021. 
	Scaled yearly sunspot numbers are depicted as a shaded area. 
	Bottom: Annual coverages of the solar wind flow data.}
	\label{fig:yearlySWTypeFractions}
\end{figure}

The hourly list of solar wind flow types is very extensive but not complete, covering about 91.9\% of all hours in 1964--2021. 
The annual coverages of solar wind flow type data are given in the bottom panel of Figure \ref{fig:yearlySWTypeFractions}. 
Flow type data coverage is almost complete since 1995 when the ACE and WIND satellites started operation.
It was very good also in 1965--1970 and 1973--1981, when several satellites were flying in the solar wind.
However, in 1982--1994, when solar wind was measured mostly by only one Earth-orbiting satellite (IMP-8), gaps in the flow type data cover almost one third of time.  
Note also that, due to the auxiliary data, the overall coverage of solar wind flow type data is clearly better than the overall coverage of solar wind data (about 76.5\%).
Here we use an updated version of the flow type list extending from November 1963 to January 2022.
(For more details on flow types and auxiliary data, see \citeA{Richardson_GA_2012}).
%These periods are typically related to long data gaps in the OMNI2 database.
				
The top panel of Figure \ref{fig:yearlySWTypeFractions} shows the annual fractions of the three solar wind types. 
Here the fractions are calculated as a ratio to all available flow type values in the respective year. 
Thus, they exclude the data gaps, and could, therefore, also be called relative flow type fractions.
The overall (relative) coverage of CME fraction is about 18.0\%.
The top panel of Figure \ref{fig:yearlySWTypeFractions} shows that the CME flow fraction varies closely in phase with the sunspot cycle.
The correlation coefficient between yearly sunspots and CME fractions is excellent, cc = 0.88, with an almost vanishing p-value of $1.7*10^{-19}$.
Even the overall level of CME fraction during each cycle closely follows the changing height of sunspot cycles, although the exact timing of cycle peaks of CME fractions may differ from sunspot maxima by 1--2 years.
The good agreement between CME fractions and sunspot activity is also seen, e.g., in the considerably lower CME fraction during the low sunspot cycle 24, compared to the level of CME fractions during the earlier, higher sunspot cycles.

% FIGURE2

The HSS/CIR flow fraction maximizes in the declining phase of the sunspot cycle, when solar polar coronal holes extend to low latitudes.
This timing difference leads to a significant negative correlation between yearly sunspots and HSS/CIR fractions (cc = -0.42; p = 0.0012).
% \remove{This is the time when solar polar coronal holes extend to low latitudes.}
Note that the (relative) HSS/CIR fraction is, on an average, 41.3\%, i.e., twice larger than the average CME fraction. 
In fact, the HSS/CIR fraction is larger than the CME fraction in almost all years except for a few sunspot maximum years. 
Interestingly, the cycle maxima of the HSS/CIR fraction seem to shift earlier in the sunspot cycle during the time interval included in Fig. \ref{fig:yearlySWTypeFractions}.
The HSS/CIR maxima are found in the pre-minimum to late declining phase in SC20--SC22, but in the early to mid-declining phase in SC23--SC24. 
Frequent high-speed streams in 1974 were found to come from persistent polar coronal hole extensions \cite{Gosling_etal_76}, while in 2003 most HSSs originated from low-latitude coronal holes \cite{Gibson_2009,Fujiki_2016, Hamada_SP_2021}. 
These differences reflect the systematic long-term  evolution in the structure of solar magnetic fields (to be discussed later in more detail).
Note also that, similarly to the CME fraction, the HSS/CIR fraction is smaller during SC24 than in earlier cycles.

The slow wind fraction depicts a small solar cycle variation with maxima typically during or soon after sunspot minima, and minima around sunspot maxima. 
Accordingly, there is a significant negative correlation between yearly sunspots and slow wind fractions (cc = -0.59; p = $1.1*10^{-6}$).
However, the most prominent feature in the slow wind fraction is its recent increase. 
Until 2002 the average slow wind fraction was 36.0\%, twice larger than the CME fraction but clearly smaller than the HSS/CIR fraction. 
However, after 2003 the average slow wind fraction is 50.5\%, making the slow solar wind the most common solar wind flow type in the last two decades. 
This recent increase in slow wind fraction during weakening sunspot activity contributes to the above-mentioned negative correlation between the two parameters.
The two highest peaks in the slow wind fraction occur around the two sunspot minima in 2009 and 2019, and attain values slightly above and below 70\%, respectively. 
This increase in the slow wind fraction during the last two decades reflects the widening of the streamer belt due to the weakening of solar polar magnetic fields \cite{Smith_Balogh_2008,Smith_chapter_2008, Wang_2009}. 
Polar fields, again, are slowly weakening since cycle 22 as a consequence of the slow decrease in the overall solar activity that characterizes the whole space age.

\section{Storm identification}
\label{Sec: Sec4 Storm identification}

We use here a fairly similar procedure to identify geomagnetic storms as adopted earlier by \citeA{Yakovchouk_2012}. 
We apply this procedure here separately both to the Dxt index and to the Dst index in order to study how the storms of different classes differ between these two indices. 						

According to the storm identification procedure, we first locate all local minima in the Dxt/Dst index in 1964--2021. 
A storm is then identified as the deepest index minimum within a 2-day (48-hour) interval from any other minimum. 
Accordingly, all minima within a 2-day interval are counted to belong to the same storm as separate (smaller) intensifications. 
This definition correctly joins together to one storm, e.g., the two possible intensifications of a CME storm, one due to the sheath and the other due to the core (ejecta), which typically have a smaller time separation of about 12--24 hours.
However, using this definition we miss one storm in those cases where two separate CMEs, each of which exceeds the storm threshold, follow each other within two days. 
While a more detailed estimate will be made in a separate study later, we have made a preliminary estimate that this leads to an underestimate of storms by less than about 10--15\%.
This is too small to have an effect on our main results and, therefore, no change due to possibly omitted storms is made here.
However, we note that a shorter time interval of 2 days was applied here, in difference to the 3-day interval used by \citeA{Yakovchouk_2012}, because the shorter separation alleviates the problem of possibly omitted storms. 
This also leads to somewhat higher numbers of storms here than in \citeA{Yakovchouk_2012}.													

\begin{figure}[h!]
	\centering
	\includegraphics[width=1\linewidth]{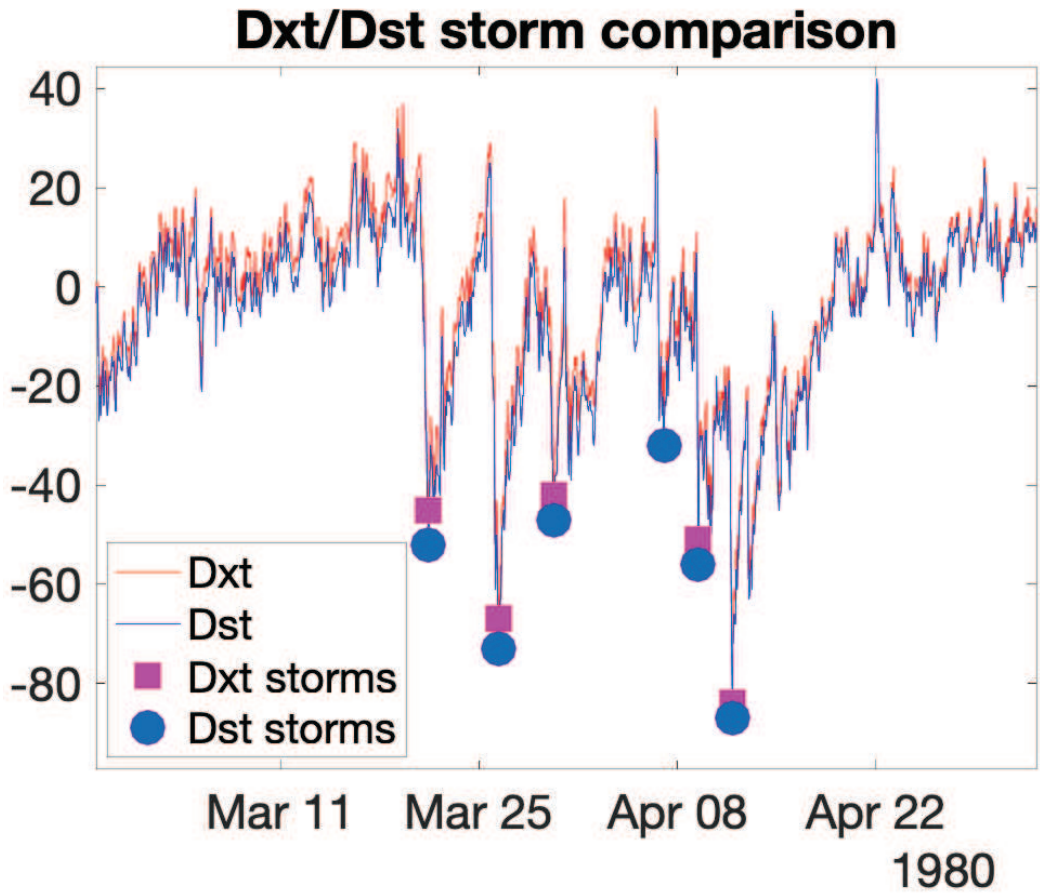}
	\caption{Dxt index (red line) and Dst index (blue line) in March--April 1980, as well as storms identified by the Dxt index (magenta dots) and the Dst index (blue dots).}
	\label{fig:Dxt_Dst_storms_comparison_04_1980}
\end{figure}

Figure \ref{fig:Dxt_Dst_storms_comparison_04_1980} depicts the Dxt index (red line) and the Dst index (blue line) in March-April 1980, as well as the storms identified by the Dxt index (magenta squares) and the Dst index (blue dots). 
The Dst index identifies six storms during this time interval, but the Dxt index only five storms. 
As discussed above, the Dxt index is, on an average, slightly higher than the Dst index and misses the weak storm in April 7 (min Dxt = -24 nT; min Dst = -32 nT). 
Overall, the selected index minima are quite appropriate to denote the peaks of storm main phases.   

% FIGURE3

We have classified the storms identified by the Dxt index (the Dxt storms) and, separately, by the Dst index (the Dst storms) according to the minimum value of the respective index into  four storm classes or categories using the following definition:												
\begin{enumerate}
\item Weak:  -50 nT $<$  Dxt, Dst $\leq$ -30 nT
\item Moderate: -100 nT $<$ Dxt, Dst $\leq$ -50 nT
\item Intense: -200 nT $<$ Dxt, Dst $\leq$ -100 nT
\item Major: Dxt, Dst $\leq$ -200 nT.
\end{enumerate}

\section{Total storm numbers during space age}
\label{Sec: Sec5 Storm numbers}

Table \ref{Table:Storms_1957} shows the storm numbers in 1957--2021 both in total and when classified into the four intensity categories, separately for the two indices. 
There are in total 2526 geomagnetic storms according to the Dxt index and 2743 storms, i.e., some 8.6\% more, according to the Dst index. 
When dividing these by the number of years (65), one finds that there have been 39/42 storms per year according to the Dxt/Dst index.
Thus, as a rule of thumb, one can say that there have been, on an average, three storms per solar rotation during the space age.

The largest category of storms, about 45\% of all storms are weak storms, while 40\% are moderate storms, 12\% intense storms and 3\% major storms.
Almost the same percentages are found for both indices, which gives evidence for the robustness of these results.	
So, roughly speaking, almost a half of the storms were weak storms and three quarters of the rest were moderate storms.
% Based on storms of the sample interval of Figure \ref{fig:Dxt_Dst_storms_comparison_04_1980} one would perhaps expect that the largest relative difference in the storm numbers between the two indices would be in the number of weak storms. 
% However, this is not the case. 
As seen in Table \ref{Table:Storms_1957}, 
There are 8.0\% more of weak storms, 10.1\% more of moderate storms, 6.9\% more of intense storms, and 4.4\% more of major storms in the Dst index than in the Dxt index. 

Table \ref{Table:Storms_1957} also lists the means of the storm minimum Dxt/Dst (storm peak) values (mean intensities), for all storms and separately for the four intensity categories. 
One can see that the two indices give very closely similar storm peak mean values of about -38\,nT, -68\,nT, -131\,nT and -277/-276\,nT for the weak, moderate, intense and major storms.
Even though the Dst index produces more storms than the Dxt index, the distribution of storm numbers as a function of storm intensity is very similar for the two indices. 
Therefore the storm mean intensities also remain quite similar.
The mean storm peak value for all storms is almost the same as for the moderate storms.

% TABLE1
\begin{table}
	\centering
	\caption{Storm numbers (with respective percentages) and mean intensities for four storm classes in 1957--2021.}
	\label{Table:Storms_1957}
\begin{tabular}{l|l|l|l|l|l}
     & Weak  & Moderate & Intense & Major & All \\
Dxt storms & 1142 (45.2\%) & 1012 (40.1\%) & 304 (12.0\%) & 68 (2.7\%) & 2526 \\
Dst storms & 1233 (45.0\%) & 1114 (40.6\%) & 325 (11.8\%) & 71 (2.6\%) & 2743 \\
 & & & & &  \\
Mean Dxt-min & -38.1\,nT & -67.5\,nT & -130.5\,nT & -277.4\,nT & -67.5\,nT \\
Mean Dst-min & -38.3\,nT & -67.7\,nT & -130.6\,nT & -275.8\,nT & -67.3\,nT \\
\end{tabular}
\end{table}

\section{Yearly storm numbers during space age}
\label{Sec: Sec6 Yearly numbers}

\begin{figure}[h!]
	\centering
	\includegraphics[width=1\linewidth]{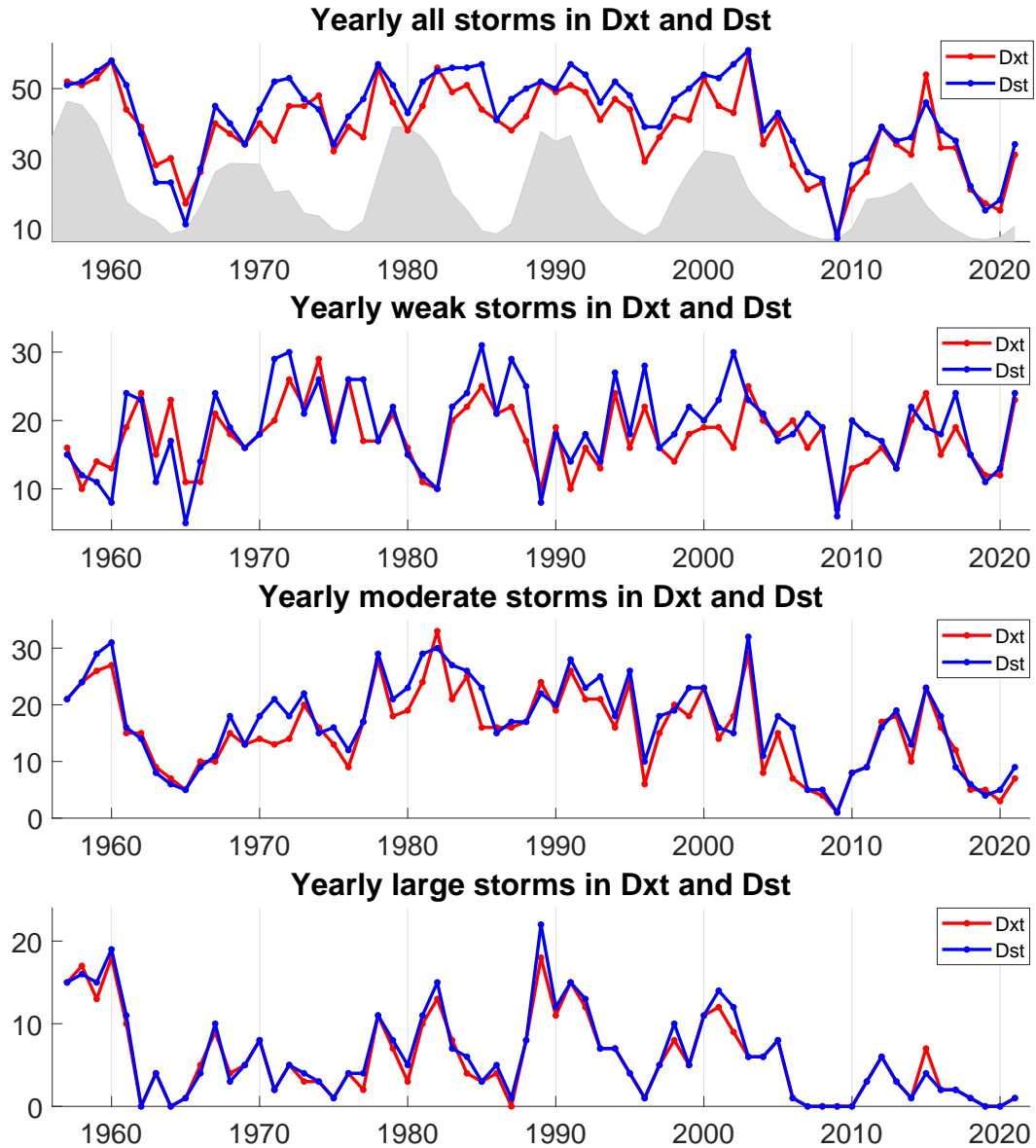}
	\caption{Yearly numbers of Dxt and Dst storms. Top: all storms with yearly sunspots depicted as gray background; Second: weak storms; Third: moderate storms; Bottom: yearly sums of intense and major storms.}
	\label{fig:yearly_storm_numbers_4intens}
\end{figure}

Figure \ref{fig:yearly_storm_numbers_4intens} depicts the yearly numbers of Dxt storms (red lines and dots) and Dst storms (blue lines and dots) during the space age. 
The top panel of Fig. \ref{fig:yearly_storm_numbers_4intens} shows the yearly numbers of all storms, with yearly sunspots depicted as gray background to indicate sunspot cycles. 
One can see that in most years the number of Dst storms is slightly larger than Dxt storms. 
Only in 8 years the Dxt index includes more storms than the Dst index. 

% FIGURE4

Despite these differences in the total number of storms, the two indices agree on the two years when the number of storms is largest: 1960 and 2003, as well as on the two years when it is smallest: 1965 and 2009. 
The two indices also depict almost the same number of storms in these four extreme years. 
The storm numbers experienced a dramatic decline in the declining phase of SC23 from 60/61 storms in 2003 to 7/8 storms in 2009. 
A similar but slightly less dramatic decline occurred in the declining phase of SC19 from 58/58 storms in 1960 to 11/17 storms in 1965.			

% During the relatively quiet years after the two years (1960 and 2003) with a maximum number of storms, the yearly Dxt and Dst storm numbers agree quite well with each other. 
After 1960, the two storm number series agree quite well with each other until the declining phase of SC20. 
The differences between Dxt and Dst storm numbers are particularly large in the declining phases of cycles 20 and 21 and around the maxima of cycles 22 and 23.
They agree very well again during the last two, rather quiet decades since 2003. 
The larger differences in yearly storm numbers during the more active cycles correspond well to the time evolution of the Dxt-Dst difference depicted in the bottom panel of Figure \ref{fig:Dxt_Dst_hourly_yearly}.

%We know that the last two decades are marked with much weaker solar and geomagnetic activity than earlier [Many Ref. Smith, Mursula ApJ].													
%Note that during those three sunspot cycles 19, 23 and 24 (and 25) when the Dxt and Dst storm numbers agree best, the storm numbers have their outstanding cycle  maximum in one year, always in the early to mid-declining phase of the cycle. However, in the other cycles 20--22, there are several years when storm numbers are almost equally large in each index, leading to a greater likelihood of the two indices depicting cycle maxima in slightly different years.													

%In cycle 20, both indices have the storm number maximum in the declining phase, but separated by two years. In cycle 21 considerable storminess was found already in the pre-maximum year 1978, as well as in the declining phase. Both indices depict two peaks with equal heights, but the declining phase maxima are again separated by a few years. In cycle 22, the Dxt and Dst storm numbers maximize in the two different sunspot number peak years of 1989 and 1991. Accordingly, the differences between Dxt and Dst storm numbers are particularly large in the declining phases of cycles 20 and 21 and around the maxima of cycles 22 and 23.

The second panel of Fig. \ref{fig:yearly_storm_numbers_4intens} shows the yearly numbers of weak storms during the space age. 
Weak storms follow a fairly systematic solar cycle variation with maxima in the declining phase of the cycle. 
During SC19--SC22 these maxima are located in the late declining phase or even at sunspot minimum, but during the last two cycles SC23--SC24 they have shifted to the early to mid-declining phase of the cycle. 
We will study this shift later in this paper when discussing the long-term evolution of solar magnetic fields and storm drivers.
During the active cycles SC20--SC22 the number of weak storms had cycle minima typically around sunspot maxima, while during SC19 and the last two cycles the weak storm minima were around sunspot minima. 
%Note also that the cycle minimum numbers of weak storms vary far less than the corresponding numbers of all storms (see top panel). 
Years 1965 and 2009, which are exceptionally low in the number of all storms (see top panel of Fig. \ref{fig:yearly_storm_numbers_4intens}), are the lowest also in weak storms but do not stand out so dramatically among other cycle minima of weak storms.
Moreover, the number of weak storms has been at a quite high level even during the weak cycle 24, around the minimum thereafter and in the first years of cycle 25.  													

The third panel of Fig. \ref{fig:yearly_storm_numbers_4intens} shows the yearly numbers of moderate storms, which have a fairly similar long-term evolution as the number of all storms. 
Here the minimum of 2009 also stands out quite dramatically.
Cycle maxima of moderate storms are mostly found slightly earlier in the declining phase than for weak storms. 
We will discuss this difference in more detail later in the connection with storm drivers.
Note that, excluding some years like 2003, there is a rather steady long-term decline in the number of moderate storms since the maximum in SC21. 
The latest minimum between SC24 and SC25 also remained quite low in the number of moderate storms.
%, relatively lower than the corresponding minimum of weak storms.

The bottom panel of Fig. \ref{fig:yearly_storm_numbers_4intens} shows the yearly numbers of intense and major storms combined together (to be called here large storms), because the yearly number of major storms is very low and vulnerable to be dominated by random fluctuations. 
The cyclic evolution of large storms, as well as their cycle maximum numbers, follow the sunspot cycles and their heights very well.
This is due to the fact that most large storms are caused by CMEs, which follow sunspots \cite<see, e.g.,>[]{Webb_and_Howard_1994, Gopalswamy_2004, Cremades_2007, Robbrecht_ASR_2006, Webb_LRSP_2012}. 
Note also the steady decline in the cycle maxima of large storms since the maximum of SC22 in 1989. 
Moreover, the 4-year period in 2007--2010 with no large storms is unique in the 65-year time interval, masking out even the latest minimum with two years (2019--2020) devoid of large storms.

% The largest differences between all yearly Dxt and Dst storm numbers (top panel) are mainly due to the corresponding differences in the number of weak storms. 
Figure \ref{fig:yearly_storm_numbers_4intens} also gives information on the differences between Dxt and Dst storm numbers.
There are 36 years when the number of weak Dst storms is larger than the number of weak Dxt storms. 
However, there are also 19 years when the number of weak Dxt storms is larger than the number of weak Dst storms. 
This leads to the perhaps somewhat surprising result that the relative difference between the Dst and Dxt storm numbers is not in weak storms but in moderate storms. 
% Based on storms of the sample interval of Figure \ref{fig:Dxt_Dst_storms_comparison_04_1980} one would perhaps expect that the largest relative difference in the storm numbers between the two indices would be in the number of weak storms. 
% The latter years reduce the contribution of weak storms to the dominance of Dst storms over Dxt storms smaller than the contribution of moderate storms.
%, where the corresponding ratio is 13/41. 
Note also that the large storm numbers are the same for Dxt and Dst in 41 years.
%, and the largest yearly difference is only 4 large storms.													

\section{Driver contributions to storm numbers}
\label{Sec: Sec7 Storm drivers}

We have identified the solar wind driver (flow type) for each storm in 1964--2021 (when there is solar wind flow type data available) by comparing the time of the storm peak (Dxt/Dst minimum) with the corresponding time in the list of solar wind flow types. 
A storm is assigned as a CME, HSS/CIR or slow wind storm, if the storm peak was inside the corresponding solar wind stream. 
Since, as discussed above, Richardson's classification has gaps, some storms remain unclassified. 
We call the latter no-SW storms. 
(Corresponding hours are called “unclear” in Richardson list.)
%\citeA{Asikainen_2016} have suggested that, very likely, most of of these hours belong to HSS/CIR streams).

The number of storms for each intensity category and solar wind driver, separately for Dxt and Dst storms, are summarized in Table \ref{Table:Storms_classified_1964}. 
Out of a total of 2202/2416 storms in 1964--2021, 2059/2244 storms (93.5\%$/$92.9\%) could be assigned to a solar wind driver. 
This large fraction guarantees that the results on the relative fractions of storms of different solar wind drivers are representative and reliable.													
Using the storm numbers of Table \ref{Table:Storms_classified_1964} we have also calculated various fractions of storms and listed them in Table \ref{Table:Fractions} and Table \ref{Table:Fractions_intensities}.
Table \ref{Table:Fractions} gives the fraction (in percentage) of storms of a certain intensity category and solar wind driver out of all solar wind-classified storms, separately for Dxt and Dst storms. 
Accordingly, the sum of all numbers in the four central columns (or only in the 'All' column) of Table \ref{Table:Fractions} add up to 100\%.
Note that the relative fraction of all weak storms has increased from Table \ref{Table:Storms_1957} due to the shorter time interval and the neglect of the active cycle 19 which has relatively more of moderate and large storms than in the whole time interval. 
On the other hand, Table \ref{Table:Fractions_intensities} lists the fraction (in percentage) of Dxt/Dst storms of certain intensity and driver out of all storms of certain intensity.
Thus, the sum of each column in Table \ref{Table:Fractions_intensities} is 100\%.

% TABLE2
\begin{table}
	\centering
	\caption{Dxt/Dst storm numbers in 1964--2021 classified according to storm intensity and storm driver.}
	\label{Table:Storms_classified_1964}
\begin{tabular}{l|l|l|l|l|l}

      & Weak  & Moderate & Intense & Major & All \\
CME & 162/154 & 371/377	& 204/218 & 48/51 & 785/800 \\
HSS & 591/640 & 386/452	& 35/37	& 0/0 & 1012/1129 \\
Slow & 203/237 & 56/76 & 3/2 & 0/0 & 262/315 \\
No SW & 75/98 & 63/66 & 5/8 & 0/0 & 143/172 \\
All & 1031/1129 & 876/971 & 247/265 & 48/51 & 2202/2416 \\
\end{tabular}
\end{table}

% TABLE3
\begin{table}
	\centering
	\caption{Fractions (in percentage) of Dxt/Dst storms of certain intensity and driver out of all Dxt/Dst storms in 1964--2021 covered by solar wind classification.}
	\label{Table:Fractions}
\begin{tabular}{l|l|l|l|l|l}

      & Weak  & Moderate & Intense & Major & All \\
CME & 7.9\%/6.9\% & 18.0\%/16.8\% & 9.9\%/9.7\% & 2.3\%/2.3\% & 38.3\%/35.7\% \\
HSS & 28.7\%/28.5\%	& 18.7\%/20.1\%	& 1.7\%/1.6\% & 0.0\%/0.0\%	& 49.2\%/50.3\% \\
Slow & 9.9\%/10.6\%	& 2.7\%/3.4\% & 0.1\%/0.1\%	& 0.0\%/0.0\% & 12.7\%/14.0\% \\
All & 46.5\%/46.0\%	& 39.4\%/40.3\% & 11.7\%/11.4\%	& 2.3\%/2.3\% & 100\%/100\% \\
\end{tabular}
\end{table}

% TABLE4
\begin{table}
	\centering
	\caption{Fractions (in percentage) of Dxt/Dst storms of certain intensity and driver out of all Dxt/Dst storms of certain intensity in 1964--2021.
	Accordingly, the sum of each column is 100\%.}
	\label{Table:Fractions_intensities}
\begin{tabular}{l|l|l|l|l}

      & Weak  & Moderate & Intense & Major \\
CME & 16.9\%/14.9\% & 45.6\%/41.7\% & 84.3\%/84.8\% & 100\%/100\%  \\
HSS & 61.8\%/62.1\%	& 47.5\%/49.9\%	& 14.5\%/14.4\% & 0\%/0\%  \\
Slow & 21.3\%/23.0\%  & 6.9\%/8.4\% & 1.2\%/0.8\%	& 0\%/0\%  \\
\end{tabular}
\end{table}

Tables \ref{Table:Storms_classified_1964} and \ref{Table:Fractions} show that the HSS/CIR streams produced 1012/1129 storms according to the Dxt/Dst index, almost exactly one half (49.2\%$/$50.3\%) of all solar wind-classified storms in 1964--2021.
As a rule of thumb, HSS/CIR streams produced some 1.5 storms per solar rotation.
% (roughly for a half of rotations one storm and for another half two storms). 
There were 785/800 CME storms, making a good third (38.3\%$/$35.7\%) of all solar wind-classified storms. 
This is very close to one CME storm per solar rotation, on an average. 
Finally, there were 262/315 (12.7\%$/$14.0\%) slow wind storms, roughly one slow wind storm in a couple of rotations.													

The three solar wind streams contributed very differently to the three storm intensity categories. 
This is most notable for the largest storms, with CME streams producing all 48/51 major Dxt/Dst storms and 204/218 (84.3\%/84.8\%) out of the 242/257 solar wind-classified intense storms. 
Altogether, CME streams were responsible for 87\% of all large (intense and major) solar wind classified storms, while HSS/CIR streams produced only 12\% and slow wind streams only 1\% of them. 
This solar wind driver division (87\%$/$12\%$/$1\%) is the same for the large storms of both indices. 													

The strongest storm in this 65-year interval occurred on 14 March 1989. 
It was, naturally, driven by a CME and reached the maximum intensity of -594/-589\,nT according to Dxt/Dst index. 
This storm was by far larger than all other storms, since the second strongest storm on 20 November 2003 reached the intensity of only -427/ -422\,nT, and all other storms remained above -400\,nT. 
No HSS/CIR storm reached even close to the threshold of a major storm, since the strongest HSS/CIR storm on 13 September 1993 had the intensity only of -155/-161\,nT. 
As seen in Tables \ref{Table:Storms_classified_1964} and \ref{Table:Fractions_intensities}, there were altogether 35/37 (14.5\%/14.4\%) intense HSS/CIR driven storms. 
Slow wind streams produced only 3/2 (1.2\%/0.8\%) intense storms.

\section{Driver contributions yearly}
\label{Sec: Sec8 Storm drivers Yearly}

\begin{figure}[h!]
	\centering
	\includegraphics[width=1\linewidth]{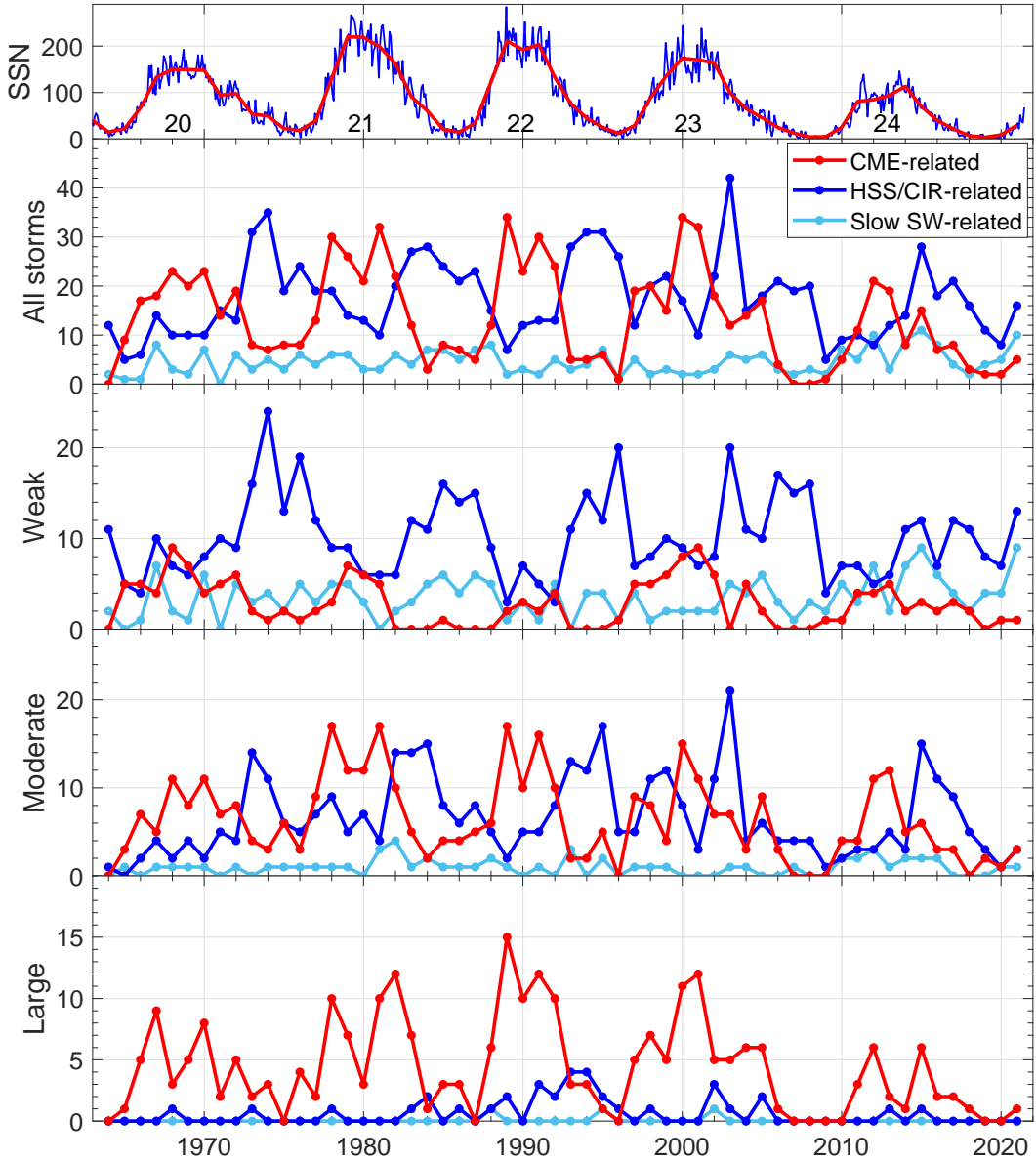}
	\caption{Top panel: Monthly (blue) and yearly (red) sunspot numbers for reference.
	Panels 2--5: Yearly storm numbers of all (panel 2), weak (panel 3), moderate (panel 4) and large (intense and major; panel 5) storms of the Dxt index, 
	separated into CME (red), HSS/CIR (blue), and slow solar wind (cyan) storms. }
	\label{fig:yearly_Dxt_storm_sw_numbers_all_3intens}
\end{figure}

\begin{figure}[h!]
	\centering
	\includegraphics[width=1\linewidth]{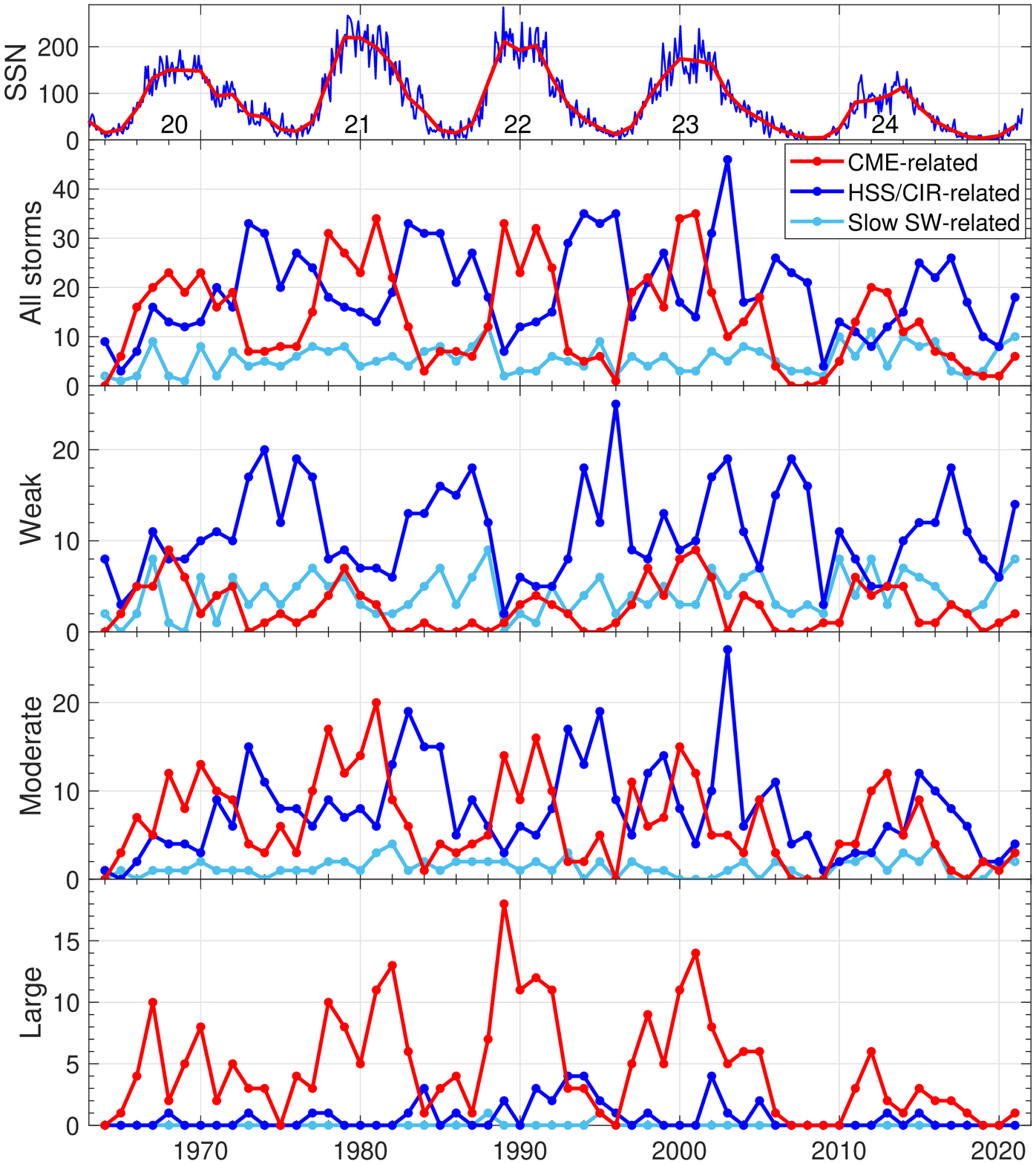}
	\caption{Same as Figure \ref{fig:yearly_Dxt_storm_sw_numbers_all_3intens}, but using storms of the Dst index.}
	\label{fig:yearly_Dst_storm_sw_numbers_all_3intens}
\end{figure}

Figures \ref{fig:yearly_Dxt_storm_sw_numbers_all_3intens} and \ref{fig:yearly_Dst_storm_sw_numbers_all_3intens} show the yearly numbers of Dxt/Dst magnetic storms of different categories in 1964--2021, separately for the three solar wind drivers: CME streams (red line and dots), HSS/CIR streams (blue) and slow solar wind streams (cyan). 
Figures \ref{fig:yearly_Dxt_storm_sw_numbers_all_3intens} and \ref{fig:yearly_Dst_storm_sw_numbers_all_3intens} depict all storms (second panel), weak storms (third panel), moderate storms (fourth panel) and large (intense and major) storms (fifth panel) related to the three drivers. 
Yearly and monthly sunspot numbers are depicted in top panel for reference.

% FIGURE5

% FIGURE6

The third panel of Figures \ref{fig:yearly_Dxt_storm_sw_numbers_all_3intens} and \ref{fig:yearly_Dst_storm_sw_numbers_all_3intens} verifies that the weak storms are mainly driven by HSS/CIR streams. 
The number of weak HSS/CIR storms greatly exceeds the number of weak CME storms or weak slow wind storms in most years, and maximizes in the declining phase of each cycle. 
Cycle maxima of the yearly number of weak HSS/CIR storms vary from 24 in SC20 to 12 in SC24.
% Even during solar cycle 24 (SC24), the maximum is quite high, although somewhat smaller than in other cycles. 
There is an interesting long-term shift in the occurrence of weak HSS storms and in the location of their cycle maxima. 
While during SC20, SC21 and SC22 most weak HSS/CIR storms are found 
% around sunspot minima, with cycle maxima 
in the late declining to minimum phase, 
%only 0--2 years before the sunspot minimum, 
during SC23 and SC24 they occur clearly earlier in the cycle, with cycle maxima in the early to mid-declining phase. 
This shift agrees with a similar shift in the location of cycle maxima of all weak storms (see Fig. \ref{fig:yearly_storm_numbers_4intens}), and proves that this shift is related to the evolution of HSS/CIR storms.
In fact, a similar shift is also seen in the fraction of HSS/CIR streams depicted in Figure \ref{fig:yearlySWTypeFractions}.
On the other hand, in accordance with the close relation between sunspots and CMEs \cite<see, e.g.,>[]{Webb_and_Howard_1994, Gopalswamy_2004, Cremades_2007, Robbrecht_ASR_2006, Webb_LRSP_2012}, the number of weak CME storms also maximizes in sunspot maximum years, but remains mostly below the number of weak HSS/CIR storms even then.

As shown in Table \ref{Table:Storms_classified_1964}, slow wind streams produce more weak storms than CME streams.
In fact, as seen in Figures \ref{fig:yearly_Dxt_storm_sw_numbers_all_3intens} and \ref{fig:yearly_Dst_storm_sw_numbers_all_3intens}, the yearly numbers of weak slow wind storms even surpass the corresponding numbers of weak HSS/CIR storms in a couple of years. 
The number of weak slow wind storms has notably increased during cycle 24. 
The average annual number of weak slow wind storms since 2010 is some 70\% larger than before 2000. 
This increase is in agreement with the increasing fraction of slow wind streams in solar wind seen in Figure \ref{fig:yearlySWTypeFractions}.

The occurrence of large CME storms (see bottom panel of Figs. \ref{fig:yearly_Dxt_storm_sw_numbers_all_3intens} and \ref{fig:yearly_Dst_storm_sw_numbers_all_3intens}) follows closely the sunspot cycle. 
Note, however, that the number of large CME storms decreases in SC24 relatively faster than, e.g., the number of weak or moderate CME storms.
As discussed above (see Table \ref{Table:Storms_classified_1964}), only 12\% of large storms are HSS/CIR storms.
Interestingly, the number of large HSS/CIR storms maximized in the declining phase of cycle 22, in early to mid-1990s, when their number even exceeded the number of large CME storms in four successive years (1993--1996). 
In fact, about 43\% of all large HSS/CIR storms occurred in six years from 1991--1996. 
This is very similarly reproduced in both Dxt and Dst storms. 
We will discuss this development in Section \ref{Sec: Sec10 HSS/CIR storms} in more detail.

Moderate storms (fourth panel in Figs. \ref{fig:yearly_Dxt_storm_sw_numbers_all_3intens} and \ref{fig:yearly_Dst_storm_sw_numbers_all_3intens}) are caused roughly equally by CME and HSS/CIR streams. 
In fact, as shown in Table \ref{Table:Storms_classified_1964}, out of the 813/905 moderate Dxt/Dst storms with solar wind data, 371/377 are driven by CMEs and 386/452 by HSS/CIRs. 
Moderate CME storms peak around the sunspot maximum, while moderate HSS/CIR storms maximize in the declining phase.
As in the case of weak HSS/CIR storms, moderate HSS/CIR storms peak somewhat earlier in the declining phase in the two recent, less active solar cycles (SC23 and SC24) than in the earlier, more active cycles.
The all-time maximum number of moderate HSS/CIR storms occurred in 2003, in the mid-declining phase of cycle 23. 
This year also marks the all-time maximum of all HSS/CIR storms during the 65-year time interval, as seen in the second panel of Figures \ref{fig:yearly_Dxt_storm_sw_numbers_all_3intens} and \ref{fig:yearly_Dst_storm_sw_numbers_all_3intens}. 
This year was the third most frequent year in high-speed streams (see Fig. \ref{fig:yearlySWTypeFractions}) that mostly emanated from persistent, isolated low-latitude coronal holes \cite{Gibson_2009,Mursula_GRL_2017,Hamada_SP_2021}.

\section{CME storms during space age}
\label{Sec: Sec9 CME storms}

We have studied magnetic storms during the space age, which is characterized by the decline of the Modern Grand Maximum when solar activity has been slowly but unsteadily decreasing.
%Figure \ref{fig:yearly_storm_numbers_4intens} shows the yearly storm numbers in three intensity classes since 1957.
As discussed above, large storms (bottom panel in Figs. \ref{fig:yearly_Dxt_storm_sw_numbers_all_3intens} and \ref{fig:yearly_Dst_storm_sw_numbers_all_3intens}) are almost exclusively CME storms, whose occurrence is related to the new magnetic flux emerging on the solar surface.
Accordingly, the temporal evolution of the yearly numbers of large storms closely follows the corresponding evolution of sunspot numbers.
% (gray background in top panel of Fig. \ref{fig:yearly_storm_numbers_4intens}).
Figure \ref{fig:yearly_storm_numbers_4intens} shows that the first years 1957--1960 include the largest average number of large storms in any sequence of four consecutive years since 1957.
This agrees very well with the fact these are the years of the maximum and early declining phase of SC19, and form the stormy peak of the MGM.

The yearly numbers of large storms follow not only the sunspot cycle but also the longer-term evolution of sunspot activity during the whole space age.
Large storms are, similarly to sunspots, greatly depleted from SC19 to SC20.
Then their number increases until SC22 and, thereafter, systematically decreases during cycles 23 and 24.
The variation of sunspot cycle amplitudes (see top panel of Fig. \ref{fig:yearly_storm_numbers_4intens}) roughly corresponds to the variation of cycle maxima of large storms, with the exception that the peak of SC22 is slightly higher in large storms than SC21, while the two cycles are roughly equally high in sunspots.
% This may reflect a weak tendency for flux accumulation, or it may be purely random.

In order to better quantify the relation between large storms and sunspots, we have calculated the correlation between yearly sunspot numbers and yearly numbers of large storms using different (yearly) lags between the two parameters.
The maximum correlation 
%between yearly sunspots and yearly number of large storms over the time interval depicted in Fig. \ref{fig:histo_4plots} 
is obtained at zero lag and has a correlation coefficient of 0.74 (p $< 2*10^{-10}$).
Accordingly, the relation between sunspots and large storms is extremely significant, and sunspots explain more than a half of the variability of yearly numbers of large storms.
This result supports the intimate connection between the two parameters.
A much higher correlation between sunspots and large storms is not even expected since, in addition to sunspots, also other, much longer-lived active regions and filaments can produce CMEs \cite{Subramanian_2001, Webb_LRSP_2012}.
% CMEs originate also outside active regions, from helmet streamers and in association with quiet Sun filament disappearances (e.g., Subramanian and Dere 2001; Webb and Howard 2012; Liu et al. 2016) and hence, CMEs are observed in all phases of the solar activity cycle. 
We also note that the correlation between sunspots and large storms is considerably reduced already at one-year lag.
% although still remains extremely significant (cc = 0.62, p $< 5*10^{-7}$), due to the considerable autocorrelation of annual sunspots.
% Thus, if there is any flux accumulation, it cannot remain effective for longer than one year.
% This reduction in lagged correlation shows that large CME storms are mainly produced by magnetic fields that have emerged on the solar surface quite recently.

Assuming that the fractions of Table \ref{Table:Fractions} are valid for the whole time period of Figure \ref{fig:yearly_storm_numbers_4intens} (so, also for the first 7 years 1957--1963 not covered by solar wind flow type data), a slightly smaller fraction than one half of all moderate storms are CME storms.
On the other hand, almost exactly one half of moderate storms are HSS/CIR storms.
As discussed above, the moderate HSS/CIR storms have their cycle maxima in the early to mid-declining phase of the solar cycle.
Accordingly, we find that the maximum correlation between the yearly numbers of sunspots and all moderate storms is obtained at one-year delay.
This correlation (cc = 0.58) is lower than for large storms but is still extremely significant (p $< 5*10^{-6}$).

%\subsection{CME storms in the declining phase of GMM}
% \subsubsection{Large CME storms}
\subsection{Large CME storms in five separate solar cycles}

% FIGURE7
\begin{figure}[h!]
	\centering
	\includegraphics[width=1\linewidth]{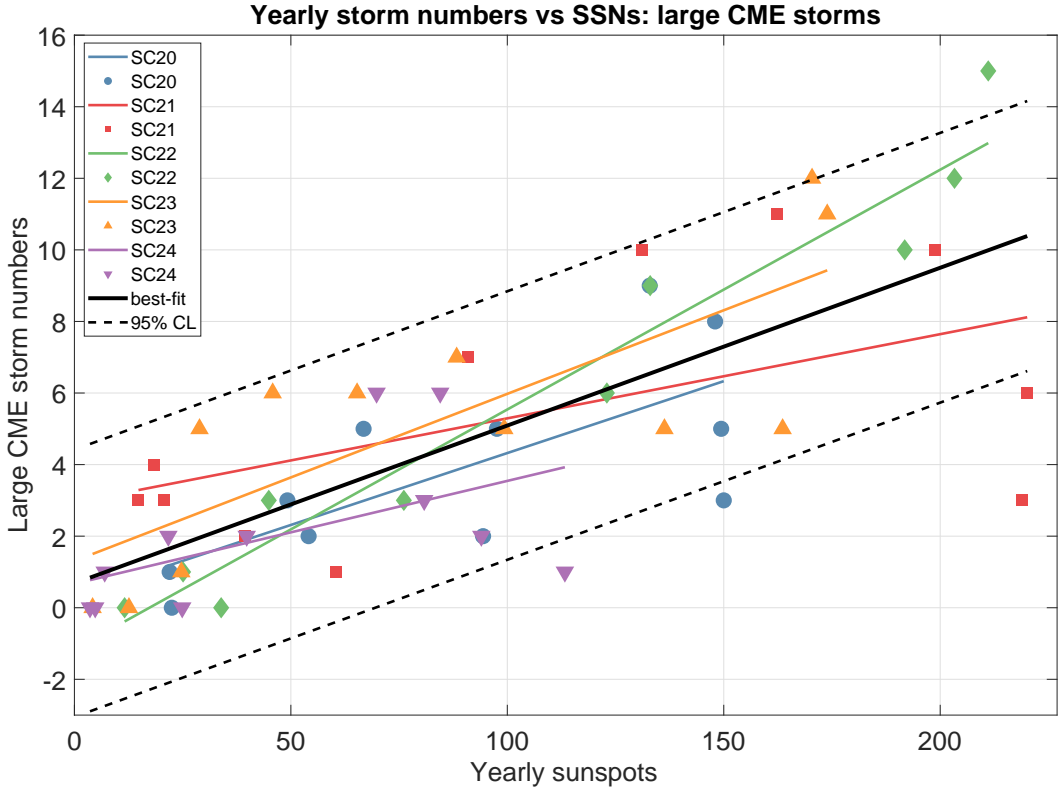}
	\caption{Scatter plot of yearly sunspots and yearly numbers of large CME storms in 1965--2019, together with the corresponding best fit line (black thick line) and the two 95\% CL lines (black dashed lines). 
	Years of the five sunspot cycles and their separate best fit lines with sunspots are marked with separate colors and markers as indicated in legend.}
	\label{fig:scatter_5SCs_yearly_SSN_storms_large_CME_bestfit_lines}
\end{figure}

% TABLE5
\begin{table}
	\centering
	\caption{Slopes, intercepts, correlation coefficients and p-values for the correlation between yearly sunspots and large CME storm numbers for five solar cycles (years in second column) and for all years in 1965--2019.}
	\label{Table:SC_bestfitlines}
\begin{tabular}{l|l|l|l|l|l}

Cycle & Years & slope & intercept & cc & p  \\
SC20 & 1965--1975 & 0.040 & 0.30 & 0.72 & 0.013 \\
SC21 & 1976--1986 & 0.024 & 2.94 & 0.55 & 0.082 \\
SC22 & 1987--1996 & 0.067 & -1.16 & 0.97 & 2.3$*10^{-6}$\\
SC23 & 1997--2008 & 0.047 & 1.31 & 0.79 & 0.002\\
SC24 & 2009--2019 & 0.029 & 0.67 & 0.53 & 0.092 \\
All & 1965--2019 & 0.044 & 0.68 & 0.76 & 2.3$*10^{-9}$ \\
\end{tabular}
\end{table}

% We can purify the sample of large storms to a sample of large CME storms by using the somewhat shorter time interval (1964--2021) that is covered with solar wind flow type data (see Figures 
Using data from the time interval (1964--2021) that is covered with information on solar wind flow types, 
%(see Figures \ref{fig:yearly_Dxt_storm_sw_numbers_all_3intens} and \ref{fig:yearly_Dst_storm_sw_numbers_all_3intens}).
% In agreement with the result of Table \ref{Table:Fractions} that only 15\% of large storms are other than CME storms, 
we find a slightly higher correlation coefficient (cc = 0.75) between the yearly numbers of sunspots and the (flow type certified) large CME storms. 
However, due to the slightly smaller number of years, the significance is slightly reduced but remains extremely high (p $<4*10^{-10}$).
In view of conducting here a study of individual, complete solar cycles, we limit the time to 1965--2019, from the start year of SC20 to the end year of SC24. 
Calculating the same correlation for these years (1965--2019), we find a similar change: the correlation coefficient is slightly increased to cc = 0.76, and the p-value is reduced to p $<2.3*10^{-9}$.
Thus, sunspots explain, on an average, about 60\% of the variability of the yearly number of large CME storms.  
We have depicted these latter values, as well as the slope and intercept of the best fit line of the corresponding linear correlation in the last row of Table \ref{Table:SC_bestfitlines}.

The decline in the yearly number of large CME storms from SC22 to SC24 depicted in the bottom panel of Figures \ref{fig:yearly_Dxt_storm_sw_numbers_all_3intens} and \ref{fig:yearly_Dst_storm_sw_numbers_all_3intens} is indeed quite dramatic and may even look exceptional compared to the corresponding decline in sunspot activity.
In order to study the relation between sunspots and large CME storms we have plotted the scatter of all yearly sunspots and yearly numbers of large CME storms in 1965-2019 in Figure \ref{fig:scatter_5SCs_yearly_SSN_storms_large_CME_bestfit_lines}.
We have included there the corresponding best fit line of linear correlation between the two parameters and the 95\% CL lines of the fit.
Note that the intercept of this best fit line is close to zero (0.68), indicating that sunspots are, statistically, prerequisite to the occurrence of large CME storms.
At intermediate sunspot activity with yearly mean of about 100, the number of large CME storms is about 5 and at higher sunspot activity of 200, twice larger.
However, the 95\% CL band is quite wide, and allows the storm numbers to vary roughly between 1 and 9 for intermediate sunspot activity and between 6 and 13 for high sunspot activity. 

We have marked in Figure \ref{fig:scatter_5SCs_yearly_SSN_storms_large_CME_bestfit_lines} the years of each of the five solar cycles (SC20--SC24) with a specific color and marker.
% Because Fig. \ref{fig:scatter_5SCs_yearly_SSN_storms_large_CME_bestfit_lines} includes only full cycles, the time interval is limited there to 1965--2019.
% The timing and length of each solar cycle were determined using the dates (year and month) of each solar minimum determined by the Solar Influences Data analysis Center (SIDC) (see http://sidc.oma.be/silso/cyclesminmax), which are based on the 13-month smoothed monthly sunspot number.
We used the sunspot minimum months given at the Solar Influences Data analysis Center (SIDC) (see http://sidc.oma.be/silso/cyclesminmax) to separate the cycles.
(The minimum years were allocated to the cycle which included a larger fraction of the respective year.)
One can see that there are "outliers" (points beyond the 95\% CL lines; stricly speaking they are not outliers but we use this word for lack of a more suitable term) in Fig. \ref{fig:scatter_5SCs_yearly_SSN_storms_large_CME_bestfit_lines} from each of the five solar cycles.
Cycle 21 has two outliers, all others one, including cycle 24, which does not deviate from the other cycles in this relation.
% cycle 22 has only one outlier, and cycles 21 and 23 four outlier years.

In order to study if there was a change in the relation between sunspots and large CME storms during SC24 or, in fact, during any of the five cycles 20--24, we have correlated yearly sunspots and large CME storm numbers separately for each of the five solar cycles.
We have plotted the corresponding five best fit lines also in Fig. \ref{fig:scatter_5SCs_yearly_SSN_storms_large_CME_bestfit_lines} using the same cycle-specific color scheme as for the yearly dots.
The slopes, intercepts, correlation coefficients and p-values of the best fit lines for the five cycles are given in Table \ref{Table:SC_bestfitlines}.
Figure \ref{fig:scatter_5SCs_yearly_SSN_storms_large_CME_bestfit_lines} and Table \ref{Table:SC_bestfitlines} show that the intercepts and even the slopes of the five best fit lines vary considerably from cycle to cycle.
The best fit lines of cycle 20 and cycle 23 have almost the same slope as the overall best fit line of all years.  
However, the intercepts of SC20 and SC23 best fit lines are quite different, leading to SC23 producing typically one large CME storm more for the same amount of sunspot activity than SC20.
Still, both of these lines remain well within the 95\% CL lines.
Correlation coefficients for SC20 and SC23 are also quite close to the correlation coefficient of the overall fit, and correlation is significant for both of these two cycles.
%although significance level remain much smaller than in the overall fit due to the much smaller number of points of correlation.

Cycles 21 and 24 have lower slopes than the slope of the overall best fit line. 
The intercept of SC21 is larger than the intercept of SC24, in fact larger than any other intercept.
Correlation coefficients for SC21 and SC24 are smaller than for other cycles, explaining only some 30\% of variation.
In fact, correlation between yearly sunspots and large CME storms is not even significant for cycles 21 and 24 at 95\% confidence level, but the fact that the respective p-values are below 0.10 indicates marginal significance.
Figure \ref{fig:scatter_5SCs_yearly_SSN_storms_large_CME_bestfit_lines} shows that SC21 has two outlier points, one of them farthest below the lower 95\% CL line.
Curiously, for both SC21 and SC24, the outlier points correspond to the highest sunspot activity years of the respective cycles.
As seen in Figures \ref{fig:yearly_Dxt_storm_sw_numbers_all_3intens} and \ref{fig:yearly_Dst_storm_sw_numbers_all_3intens}, there are two peaks in the number of large CME storms during these cycles, separated by a dramatic dropout exactly at the respective sunspot maxima (years 1979 and 1980 in SC21 and year 2014 in SC24).
This evolution, which essentially deteriorates the studied correlation in these cycles, closely reminds the structure of the Gnevyshev gap \cite{Gnevyshev_1977, Richardson_JGR_2000, Storini_2003}. 
A simultaneous Gnevyshev gap in sunspots is clearly visible (although far less strong than in storm number) only in SC24, but not in SC21 where sunspots even reach the highest yearly value among the five cycles studied.

Cycle 22 deviates from the other cycles in all aspects. 
As seen in Table \ref{Table:SC_bestfitlines}, the slope of SC22 is the largest, more than twice the slopes of SC21  and SC24, and some 50\% larger than the slope in SC20 and SC23.
Most dramatically, there is an almost perfect correlation between the yearly sunspots and large CME storms in SC22, with sunspots explaining almost 95\% of the annual variation of large CME storms.
The close relation between sunspots and large CME storms in SC22 can also be seen (see Figs. \ref{fig:yearly_Dxt_storm_sw_numbers_all_3intens} and \ref{fig:yearly_Dst_storm_sw_numbers_all_3intens}) in that both sunspots and large CME storms have SC22 maximum in 1989 and another, slightly lower peak in 1991.
% Large CME storms have a closely similar temporal variation during SC22.
This is another demonstration of a Gnevyshev gap that is simultaneous in both parameters.
%but, compared to SC21 and SC24, the gap in SC22 is narrower by one year and the reduction of storm numbers in the gap is not as large in SC22 as in the other two cycles.

\subsection{Moderate CME storms in five separate solar cycles}

% FIGURE8
\begin{figure}[h!]
	\centering
	\includegraphics[width=1\linewidth]{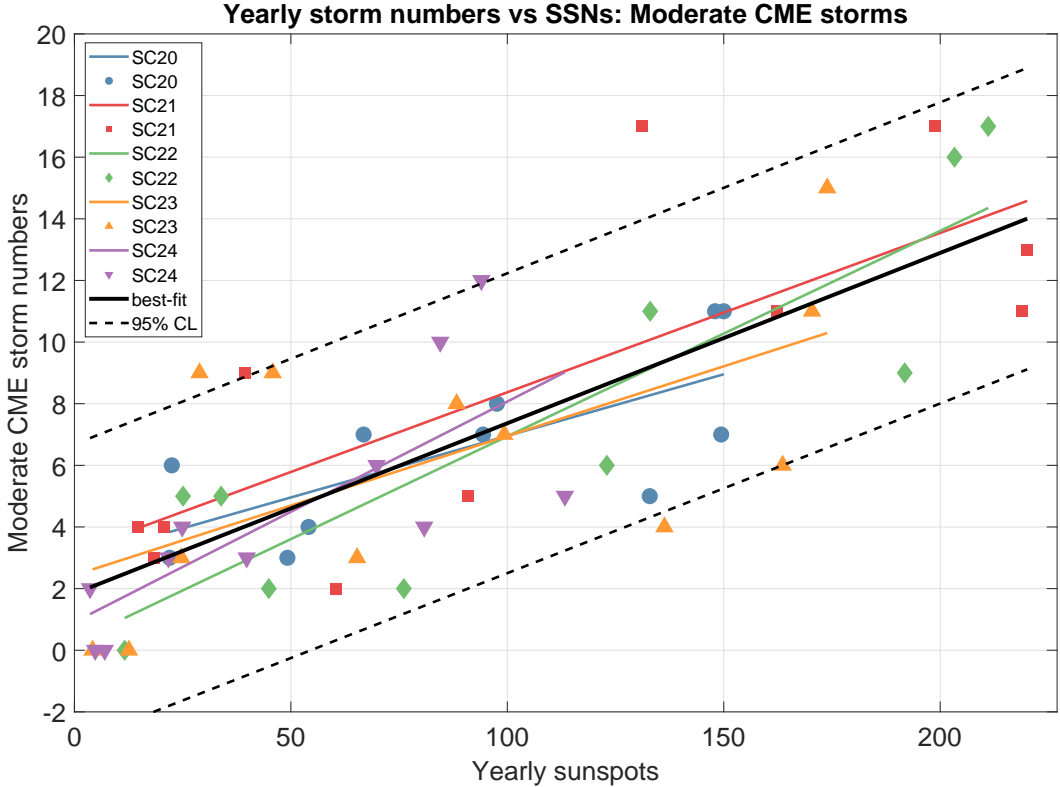}
	\caption{Same as Figure \ref{fig:scatter_5SCs_yearly_SSN_storms_large_CME_bestfit_lines} but for moderate CME storms. }
	\label{fig:scatter_5SCs_yearly_SSN_storms_moderate_CME_bestfit_lines}
\end{figure}

% TABLE6
\begin{table}
	\centering
	\caption{Slopes, intercepts, correlation coefficients and p-values for the correlation between yearly sunspots and moderate CME storm numbers for five solar cycles (years in second column) and for all years in 1965--2019.}
	\label{Table:SC_bestfitlines_moderate}
\begin{tabular}{l|l|l|l|l|l}

 Cycle & Years & slope  & intercept & cc & p  \\
SC20 & 1965--1975 & 0.040 & 2.96 & 0.72 & 0.012 \\
SC21 & 1976--1986 & 0.052 & 3.20 & 0.78 & 0.005 \\
SC22 & 1987--1996 & 0.067 & 0.27 & 0.88 & 0.0007\\
SC23 & 1997--2008 & 0.045 & 2.43 & 0.64 & 0.025\\
SC24 & 2009--2019 & 0.072 & 0.91 & 0.76 & 0.006 \\
All & 1965--2019 & 0.055 & 1.84 & 0.78 & 1.8$*10^{-12}$ \\
\end{tabular}
\end{table}

Figure \ref{fig:scatter_5SCs_yearly_SSN_storms_moderate_CME_bestfit_lines} depicts similar correlations for moderate CME storms as Figure \ref{fig:scatter_5SCs_yearly_SSN_storms_large_CME_bestfit_lines} for large CME storms.
The corresponding parameters of correlations are given in Table \ref{Table:SC_bestfitlines_moderate}.
The slope of the correlation (0.055) between sunspots and all moderate CME storms (bottom row of Table \ref{Table:SC_bestfitlines_moderate}) is larger than in the case of all large CME storms (0.044), reflecting the overall larger number of moderate CME storms (see Table \ref{Table:Storms_classified_1964} and Figs. \ref{fig:yearly_Dxt_storm_sw_numbers_all_3intens} and \ref{fig:yearly_Dst_storm_sw_numbers_all_3intens}).
The correlation coefficient for all moderate CME storms (0.78) is almost the same as for large CME storms (0.76), but the p-value is three orders of magnitude smaller, indicating improved significance due to larger statistics. 

Comparing the best fit correlation lines of the five cycles in Figure \ref{fig:scatter_5SCs_yearly_SSN_storms_moderate_CME_bestfit_lines} to the corresponding lines in Figure \ref{fig:scatter_5SCs_yearly_SSN_storms_large_CME_bestfit_lines}, one can see that they are more coherently aligned between themselves and also with the overall best fit line in Figure \ref{fig:scatter_5SCs_yearly_SSN_storms_moderate_CME_bestfit_lines}.
As Tables \ref{Table:SC_bestfitlines} and \ref{Table:SC_bestfitlines_moderate} show, the slopes vary from 0.040 to 0.072 (by 0.032) for moderate storms and from 0.024 to 0.067 (by 0.043) for large storms.
Similarly, the spread of the five intercepts is smaller for moderate than large CME storms. 
%This reflects the greatly improved significance of the overall correlation for moderate storms.
Accordingly, the improved significance of the correlation between all moderate CME storms and sunspots is due to the five cycles having more similar correlations for moderate CME storms than for large CME storms.
This is most likely due to the fact that the yearly number of moderate CME storms is larger than the number of large CME storms, which reduces statistical fluctuations between the five cycles for moderate CME storms.
Note also that 
% although the p-values of those two cycles (SC22 and SC23) that had the best correlation for large CME storms, are larger for moderate CME storms, 
the correlation between sunspots and moderate CME storms is significant for all five cycles.
% and better than for large CME storms for three cycles.
% This also affects to the higher overall correlation and is, most likely, due to improved statistics.

Curiously, for some cycles, the correlation parameters remain surprisingly similar for moderate and large CME storms.
This is most clearly valid for SC20 for which the slopes and correlation coefficients are exactly the same.
Even the p-values are almost the same. 
% Only the intercept is larger for moderate CME storms in SC20, suggesting that 
The slopes of moderate CME storms remain almost the same as for large CME storms also in SC22 and SC23, but the significance of correlation is higher for large CME storms in these two cycles.
Still, correlation is highly significant in SC22, where the correlation coefficient for moderate storms (as for large storms) is highest among all cycles, with sunspots explaining almost 80\% of the annual variation of moderate CME storms.
% However, for both two cycles the significance of correlation is smaller for moderate CME storms, but still within the 95\% confidence limit.
SC23 has the weakest correlation and the largest number of outlier points for moderate CME storms.
The slopes and correlation coefficients in SC21 and SC24 are larger for moderate than large CME storms.
Correlations are also 95\% CL significant for moderate CME storms (but only marginally significant for large CME storms) in these two cycles.
The slope for moderate CME storms is largest in SC24, even slightly larger than the corresponding slope in SC22.
% Cycle 20 had again the smallest number of outliers.

\subsection{CME storms at solar minima}

% FIGURE9   scatter_yearly_SSN_allCME_storms_3minyrs_fit_minyrs_noSC23_lines_allCMEline_NEW
\begin{figure}[h!]
	\centering
	\includegraphics[width=1\linewidth]{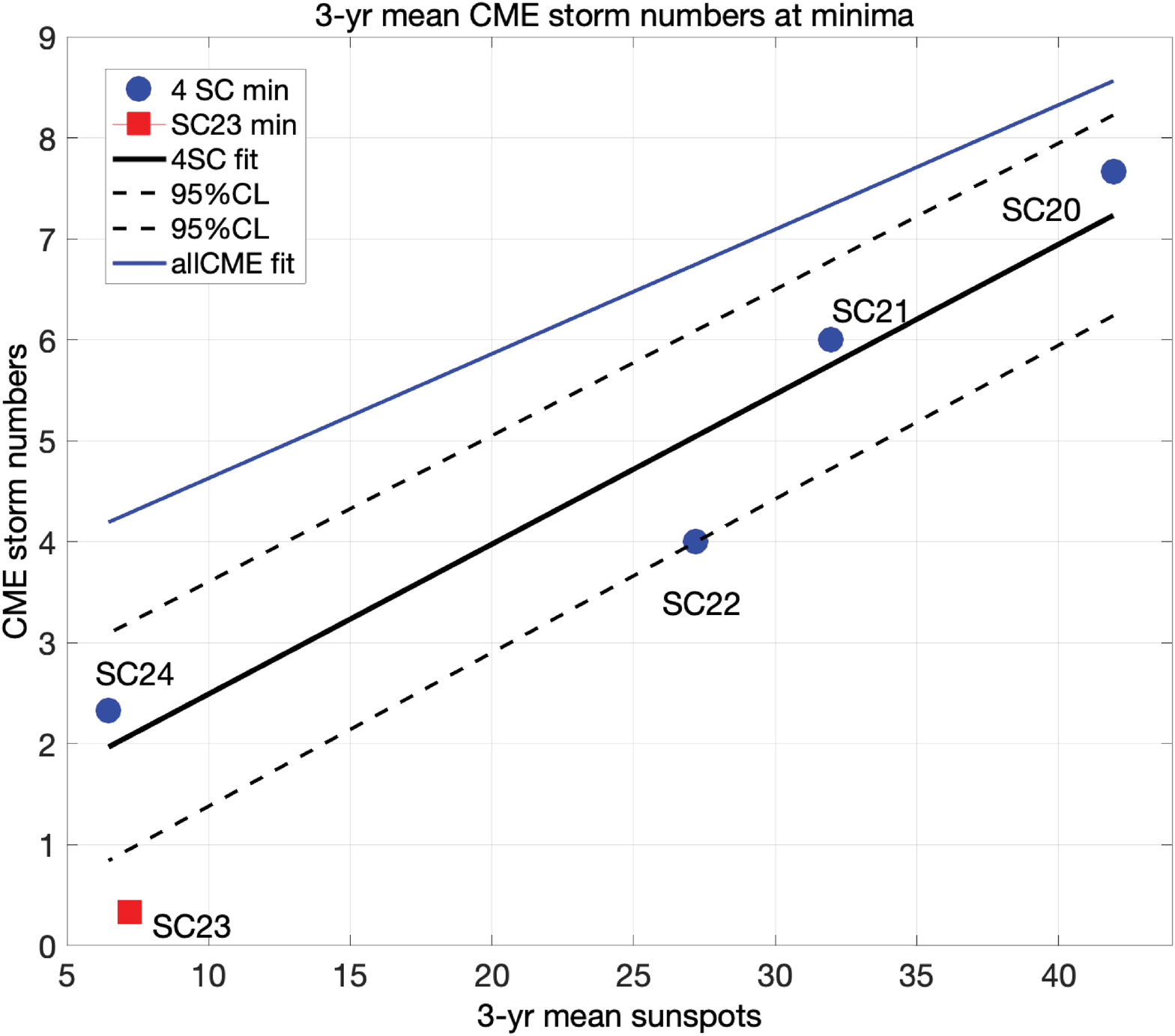}
	\caption{Scatter plot of three-year means of sunspots and CME storms calculated for the minimum years of the four solar cycles SC20--SC22 and SC24 (blue circles),
	together with the corresponding best fit line (black thick line) and the two 95\% CL lines black dashed lines).
	Red square denotes SC23 which is not included in the fit.
	Blue solid line denotes the best fit line between yearly sunspots and all CME storm numbers.
	}
	\label{fig:scatter_yearly_SSN_allCME_storms_3minyrs_fit_minyrs_noSC23_lines_allCMEline_NEW}
\end{figure}

A notable feature in the yearly number of all CME storms (see the second panel of Figs. \ref{fig:yearly_Dxt_storm_sw_numbers_all_3intens} and \ref{fig:yearly_Dst_storm_sw_numbers_all_3intens}) is the low number of CME storms around the sunspot minimum between cycles 23 and 24 (to be called minimum of SC23).
In fact, there were no CME storms at all in two successive years of 2007--2008.
There was only one other year (1964) with no CME storms within the whole 58-year time interval (1964--2021) with solar wind flow data.
Moreover, since the solar wind flow data was very partial in 1964 (see Fig. \ref{fig:yearlySWTypeFractions}), the lack of CME storms in 1964 is not as solidly founded as in 2007--2008 when the coverage was almost 100\%.
Even the latest solar minimum after the weak cycle SC24 had at least two CME storms in each year.
% Large CME storms were absent even in 2010. 
Moreover, during the four-year time interval in 2006--2009, there were altogether only 5 CME storms, out of which only one was large.
In comparison, there were 12 (17) CME storms in the second-weakest (third-weakest) 4-year minimum of 2018--2021 (1993--1996, respectively) in 1964--2021.
Note also that these results on the minimum-time storm numbers are similarly reproduced in both the Dxt and Dst indices, which gives strong support on their validity.

In order to study the relation between yearly sunspots and yearly CME storm numbers during solar minima, we calculated the mean of yearly CME storm numbers in three successive years around each solar minimum when the number of CME storms was smallest.
Figure \ref{fig:scatter_yearly_SSN_allCME_storms_3minyrs_fit_minyrs_noSC23_lines_allCMEline_NEW} shows these three-year mean CME storm numbers for the five cycle minima as a function of the corresponding three-year mean sunspot numbers.
(Cycles 20--22 and 24 are depicted with blue circles, cycle 23 is depicted as a separate red square).
We have also plotted there the best fit line (and the 95\% CL lines) for the correlation between the sunspots and the three-year CME storm numbers for the four minima (cycles 20--22 and 24), leaving the minimum of SC23 out of the fit.
The correlation coefficient is fairly high (cc = 0.95) and correlation is significant but, because of the small number of points, the p-value is only 0.047.
% The slope of the best fit line is 0.15 and intercept is 1.01.

Interestingly, Figure \ref{fig:scatter_yearly_SSN_allCME_storms_3minyrs_fit_minyrs_noSC23_lines_allCMEline_NEW} shows that the three-year mean CME activity at the minimum of SC23 is below the best fit correlation line of the four other cycles.
%and, naturally, below the best fit line for all CMEs.
This gives additional evidence for the view that CME activity in the minimum after cycle 23 was exceptionally weak, and underlines the uniqueness of this minimum among all other solar minima during the space age.

We have also included in Figure \ref{fig:scatter_yearly_SSN_allCME_storms_3minyrs_fit_minyrs_noSC23_lines_allCMEline_NEW} the best fit line for the correlation between yearly sunspots and yearly numbers of all CME storms.
This correlation is better (cc = 0.855; p $< 10^{-16}$) than between yearly sunspots and large CME storms (see Fig. \ref{fig:scatter_5SCs_yearly_SSN_storms_large_CME_bestfit_lines} and Table \ref{Table:SC_bestfitlines}) or sunspots and moderate CME storms (Fig. \ref{fig:scatter_5SCs_yearly_SSN_storms_moderate_CME_bestfit_lines} and Table \ref{Table:SC_bestfitlines_moderate}).
The best fit line of the all-CME correlation is above the 95\% confidence lines of the minimum time correlation.
However, since the slope (0.123) of the all-CME best fit line is slightly smaller than the slope for the minimum time correlation, the all-CME line will fall within the 95\% CL limit at sunspot value of about 55--60 and will cross the three-year minimum best fit line at about 95.
This indicates a small nonlinearity in the relation between sunspots and CME storms at small sunspot numbers.

\section{HSS/CIR storms and HCS evolution}
\label{Sec: Sec10 HSS/CIR storms}

% FIGURE10
\begin{figure}[t!]
	\centering
	\includegraphics[width=0.9\linewidth]{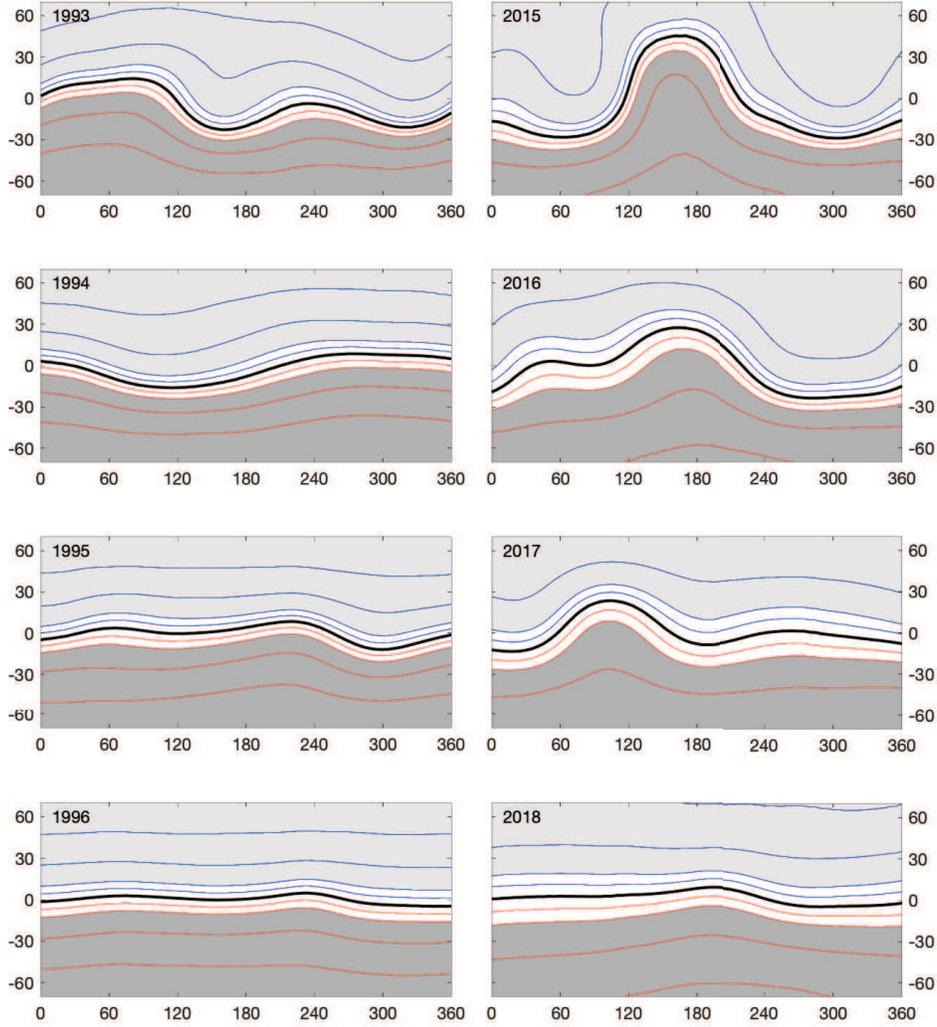}
	\caption{WSO coronal source surface synoptic maps (radial model with source surface distance at 3.25 solar radii) for one (June--July) month in (left column) 1993 (CR1870), 1994 (CR1884), 1995 (CR1897), 1996 (CR1910) and in (right column) 2015 (CR2165), 2016 (CR2178), 2017 (CR2192), 2018 (CR2205).
	Maps are redrawn using WSO data and depict the coronal source surface field for one Carrington rotation in longitude (x-axis from 0$^\circ$ to 360$^\circ$) - latitude (y-axis; from -70$^\circ$ to +70$^\circ$) grid. 
	 Light gray background color denotes the positive polarity region, dark gray color the negative polarity region. (Both minima are positive polarity times). 
	 Neutral line is marked in black thick line and corresponds to the center of the heliospheric current sheet. 
	 Curves on both sides of the neutral line (above NL: blue line; below NL: red line) mark isolines of the source surface field intensity at $\pm0.5, \pm1, \pm2.5$, and $\pm5$ $\mu$T.
%	 The horizontal green line denotes the solar equator which in June is close to the ecliptic.
%	 Carrington rotation number is denoted below each map.
}
	\label{fig:WSO_HCS_8plots}
\end{figure}

As noted in Section \ref{Sec: Sec8 Storm drivers Yearly}, about 43\% of all large HSS/CIR storms occurred in the six years from 1991--1996 (see bottom panel of Figs. \ref{fig:yearly_Dxt_storm_sw_numbers_all_3intens} and \ref{fig:yearly_Dst_storm_sw_numbers_all_3intens}). 
Such an exceptional occurrence of intense HSS storms at this time is most likely due to the fact that the HCS was exceptionally thin during the declining phase of SC22 and the subsequent minimum \cite{Richardson_1997}, allowing the Earth to reach higher heliomagnetic latitudes and more of high-speed streams than during the corresponding time of other solar cycles. 
Large HSS/CIR storms had their all-time maximum of four storms per year in two consecutive years in 1993--1994.
However, interestingly, the moderate HSS/CIR storms (see the fourth panel of Figs. \ref{fig:yearly_Dxt_storm_sw_numbers_all_3intens} and \ref{fig:yearly_Dst_storm_sw_numbers_all_3intens}) had their cycle maximum one year later, in 1995, and the weak HSS/CIR storms (the third panel of Figs. \ref{fig:yearly_Dxt_storm_sw_numbers_all_3intens} and \ref{fig:yearly_Dst_storm_sw_numbers_all_3intens}) in 1996.
(The latter was the all-time maximum for Dst and second highest peak for Dxt).

A similar temporal ordering can be found in the peak number of large, moderate and weak HSS/CIR storms during all cycles of the space age.
Moderate HSS/CIR storms tend to peak not earlier than large HSS/CIR storms, and weak HSS/CIR storms not earlier than moderate HSS/CIR storms.
During the declining phase of SC23, the peak number of large HSS/CIR storms (3/4 in Dxt/Dst) was found in 2002, while the peak of moderate and weak HSS/CIR storms was one year later in 2003.
During SC24, one large HSS/CIR storm occurred in 2013 and also in 2015, according to both indices.
Moderate HSS/CIR storms had their maximum in 2015, and weak storms in 2015 and 2017 in Dxt and only in 2017 in Dst.  

During the earlier cycles, the peak numbers and timings are less certain because of the larger number of data gaps in solar wind flow classification, especially in the 1980s (see bottom panel of Fig. \ref{fig:yearlySWTypeFractions}).
Still, the same ordering is followed during the declining phase of SC20, with the peak of large and moderate HSS/CIR storms occurring in 1973, and weak HSS/CIR storms in 1974.
% The peak number in weak HSS/CIR storms (24/20 in Dxt/Dst; see Figs. \ref{fig:yearly_Dxt_storm_sw_numbers_all_3intens} and \ref{fig:yearly_Dst_storm_sw_numbers_all_3intens}) was found in 1974, while the peak of moderate HSS/CIR storms (14/15) one year earlier in 1973. 
% Only one large HSS/CIR storm occurred in 1974.
% the number of large (intense) HSS/CIR storms are typically very small and vulnerable to statistical uncertainly.
% Cycle 21 was already a bit more active in large HSS/CIR storms which had their peak (2/3 large storms) in 1984, the same year as moderate HSS/CIR storms and one year before the peak in weak HSS/CIR storms.
In cycle 21, according to Dxt, large and moderate HSS/CIR storms had their peak in 1984 and weak storms in 1985.
According to Dst, they occurred in 1984, 1983 and 1987, respectively.
Note that the results on the temporal ordering of HSS/CIR storm numbers do not only apply to storm peaks but also to the temporal ordering of the bulk of corresponding HSS/CIR storms.
For all cycles and for both storm indices, the mean of the distribution of moderate HSS/CIR storms occurs earlier in the declining phase than the mean of weak HSS/CIR storms.
%Also, the larger the number of large HSS/CIR storms in cycle is, the better this ordering also applies to them since random fluctuations are reduced and the pattern is better held.
% increasing statistical confidence. 

This temporal ordering in the occurrence of HSS/CIR storms of different intensity is related to a systematic evolution of the heliospheric magnetic field, in particular of the structure of the heliospheric current sheet, in the declining phase of the solar cycle. 
Figure \ref{fig:WSO_HCS_8plots} depicts coronal (source surface) synoptic maps of the Wilcox Solar Observatory (WSO) for one solar rotation in four years in the mid-1990s (1993--1996), in the declining to minimum phase of SC22 (left column), as well as in four years in the late 2010s (2015--2018), in the declining phase of SC24 (right column).
We have selected a rotation including the June month of each year, because the Earth is then close to the solar equator, viewing both hemispheres roughly equally.
The coronal (source surface) magnetic field is calculated from photospheric field observations at WSO with a potential field source surface (PFSS) model \cite{Altschuler_1969, Schatten_1969, Hoeksema_1983}.
We have selected here the radial model, which assumes that the photospheric field is radial, and the distance of the source surface is at 3.25 solar radii.  
This model needs no polar correction and depicts the structure of the HCS very clearly.
% This model needs no polar correction and gives the best overall match to the structure of the inner corona and to the estimated tilt of the HCS (REF!?).

Neutral line, the solar magnetic equator and the center of the heliospheric current sheet, is denoted as a thick black line in each map of Figure \ref{fig:WSO_HCS_8plots}.
Neutral line divides the coronal (source surface) magnetic field into two opposite polarity regions, which for both SC22 and SC24 are ordered in the same sense of positive (negative) polarity field dominating in the northern (southern, respectively) hemisphere.
(Both are minima of positive solar polarity).
Positive polarity field is depicted in the synoptic maps with light gray color and negative polarity field with dark gray.
The white region between these two opposite-polarity regions is the heliospheric current sheet (streamer belt), whose center is the neutral line.
Synoptic maps also include other curves which tend to roughly follow the neutral line structure.
These curves are colored in blue above the neutral line and in red below it, 
% (black lines are inside the dark gray region and do not show very well), 
and mark the isolines of the coronal (source surface) field intensity at $\pm0.5, \pm1, \pm2.5$, and $\pm5\,\mu$T. 
The $\pm1\,\mu$T lines (second blue and second red lines) define the HCS region and their latitudinal separation is used as an approximate width of the HCS.
Accordingly, the closer the isolines are to each other, the larger the magnetic field gradients around the neutral line are, and the closer to the NL the polar coronal holes and the related high-speed streams are.

Figure \ref{fig:WSO_HCS_8plots} shows that the estimated HCS is rather narrow, about 20--25$^\circ$ wide in heliographic latitude during all the years of SC22 (1993--1996), while it is considerably wider, about 35--40$^\circ$ in SC24 (2015--2018).
This is in agreement with the above noted exceptionally thin HCS during the declining phase of SC22 and the subsequent minimum \cite{Richardson_1997}.
The thin current sheet allows a better access for the fast solar wind from polar coronal holes to low heliolatitudes.
Therefore, the Earth is, on an average, more exposed to the high-speed solar wind streams in the 1990s than, e.g., in the 2010s.
% The horizontal green line denotes the solar equator which in June is close to the ecliptic plane, and indicates the approximate location of the Earth with respect to the estimated location of the HCS.

% The tilt of the heliospheric current sheet slowly decreases during the declining phase, whence the Earth is less and less exposed to fast winds when moving from the early declining phase towards the solar minimum. 
% While this evolution is, because of the exceptionally thin HCS during the declining phase of SC22, particularly clear in the 1990s, a fairly similar, but temporally shorter evolution can be seen in the early declining phase of SC23 in 2002--2003 and, less clearly and systematically, in the declining phase of all cycles.
% Moderate HSS/CIR storms peak somewhat earlier in the declining phase than the weak HSS storms.

The occurrence and temporal duration of high-speed solar wind streams at the Earth depends not only on the thickness of the HCS, but also on its curviness, which is mainly determined by the solar dipole tilt angle. 
The dipole tilt angle has a systematic variation along the solar cycle.
It has its maximum of 90$^\circ$ at the time of polarity reversal, which happens close to the sunspot maximum.
The tilt angle decreases slowly but unsteadily during the declining phase to its minimum (typically close to 0$^\circ$) at or soon after the sunspot minimum.
% This decrease takes place during the declining phase of every solar cycle. 
One can see in Figure \ref{fig:WSO_HCS_8plots} the typical decrease of the HCS tilt (curviness) from 1993 to 1996 (left maps) and from 2015 to 2018 (right maps).
%and the same evolution takes place during the declining phase of all solar cycles.
Note that these maps are from slightly different parts of the declining phase of the two cycles, reflecting the different evolution of polar fields.
This also leads to the different timing of HSS/CIR storms during these two cycles, as discussed above.
In SC24, the first map in 2015 is from the early declining phase, while the first year of SC22 (1993) is already in the mid-declining phase.
Accordingly, the tilt is slightly larger in 2015 than in 1993 (see top row of Fig. \ref{fig:WSO_HCS_8plots}).
Anyway, in both years, the HCS is quite curved and regions of positive magnetic polarity from the northern hemisphere intrude into the southern hemisphere, and vice versa.
This is clear, e.g, in 1993 when the northern field intrudes into the southern hemisphere in longitude range of about 120--210$^\circ$ and about 270--360$^\circ$, or in 2015 when the southern field intrudes strongly into the northern hemisphere in the longitude range of about 110--220$^\circ$.
During such intrusions the Earth is subjected to the effect of HSS streams from coronal holes (and related CIRs developing during the passage from the Sun to the Earth).
Since the solar wind speed increases with heliomagnetic latitude, i.e., with the distance from the neutral line, the further the intrusion takes the neutral line from the solar equator (or, more exactly, from the ecliptic), the faster the solar wind measured at the Earth is and the longer the HSS stream lasts.
Thus, since the tilt is decreasing during the declining phase of the solar cycle, the intrusions most effectively produce large HSS/CIR storms in the early to mid-declining phase.
Moreover, since the HMF intensity has its cycle maximum in the early declining phase, the southward HMF component, which controls the reconnection in the dayside magnetosphere, is, on an average, also slightly stronger at this time.
These two effects lead to the fact that the cycle maxima of large HSS/CIR storms are typically seen in the early to mid-declining phase, before the corresponding maxima of moderate or weak HSS/CIR storms.

% These are typical HCS responses to elephant trunk extensions of polar coronal holes to low latitudes, and the Earth receives the fastest solar wind speeds from such coronal hole extensions.
% Note that the farther the Earth is from the HCS region, the faster solar wind it receives and the more likely the HSS/CIR storm will be intense or moderate.

In 1994 and in 2016 (second row of Fig. \ref{fig:WSO_HCS_8plots}), the HCS structure is quite dipolar, but the tilt is already smaller than one year earlier.
Although the Earth's largest distance from the NL is now smaller (Earth reaches slightly lower heliomagnetic latitudes) than one year earlier, the Earth is still during a large fraction of time outside of the HCS. 
The tilt is further reduced in 1995 and 2017 (third row of Fig. \ref{fig:WSO_HCS_8plots}), and again in 1996 and 2018 when the Earth was mostly within the HCS region, at least in June. 
% which is depicted in Fig. \ref{fig:WSO_HCS_8plots}.
As seen in Figures \ref{fig:yearly_Dxt_storm_sw_numbers_all_3intens} and \ref{fig:yearly_Dst_storm_sw_numbers_all_3intens}, the number of moderate HSS/CIR storms in SC24 systematically declines from the maximum in 2015 onwards, as the tilt is decreasing.
During SC22, the maximum of large HSS/CIR storms was found in 1993--1994 when the tilt was still fairly large, while the moderate HSS/CIR storms peaked in 1995.
Note that Figure \ref{fig:WSO_HCS_8plots} depicts the HCS in June when the Earth is at low heliographic latitudes.
However, during the high-latitude periods in Spring and Fall, when the Earth is below or, respectively, above the solar equator, the HCS thickness plays a crucial role.
Then, if the HCS is narrow, as in the declining phase of SC22, the Earth can be exposed to fast solar wind streams around equinoxes.
For example, in 1996, at the maximum year of weak HSS/CIR storms of SC22, some 75\% of weak storms occurred in Spring or Fall, and only 25\% in Summer and Winter.

Accordingly, the evolution of the properties (in particular thickness and tilt) of the HCS and solar magnetic fields over the solar cycle and over the whole space age can explain all the observed changes in the occurrence of HSS/CIR storms noted in earlier Sections.
First of all, the temporal ordering of the cycle maxima of large, moderate and weak HSS/CIR storms (in this order) in the declining phase of each solar cycle reflects the systematic decrease of the HCS tilt (curviness) during the respective phase of each solar cycle.
This decrease, again, follows the regular dynamics of solar magnetic fields that also control the solar dipole tilt.
As explained in the Introduction, surges of new magnetic flux create extensions of polar coronal holes to lower latitudes and, thereby, affect the dipole tilt. 
When activity subsides in the declining phase, less surges appear and the dipole tilt decreases.
Since large (weak) HSS/CIR storms require a larger (smaller, respectively) tilt, their occurrence maxima follow the decreasing tilt during the declining phase.
We note that the HCS affects the HSS/CIR storm occurrence even at other times of the solar cycle, but the effect is far less clear and systematic because the solar wind speed does not have the same latitudinal ordering at other times (especially around solar maxima) as in the declining to minimum phase of the cycle. 

Secondly, the change in overall solar activity during the space age and its effect upon the structure of solar magnetic fields and the HCS can explain the shift in the location of cycle maxima of HSS/CIR storms.
As noted in Section \ref{Sec: Sec8 Storm drivers Yearly}, during SC20--SC22 the cycle maxima of large and moderate HSS/CIR storms are located in the late declining phase, but during the last two cycles SC23--SC24 they have shifted to the early to mid-declining phase of the cycle (see two lowest panels of Figs. \ref{fig:yearly_Dxt_storm_sw_numbers_all_3intens} and \ref{fig:yearly_Dst_storm_sw_numbers_all_3intens}). 
The same shift is also seen in the timing of cycle maxima of all weak storms (see second panel of Fig. \ref{fig:yearly_storm_numbers_4intens}).
After the reversal of the solar dipole, surges of new-polarity magnetic flux strengthen the polar fields.
In the case of weak cycles, there are fewer surges, whence the polar fields remain weaker, polar coronal holes smaller and the HCS region wider.
Also, extensions of polar coronal holes only occur in the early to mid-declining phase of the cycle, leading to early maxima of HSS/CIR storms in weak cycles.
Moreover, because of the thick HCS, the Earth stays within the slow wind of the HCS region from quite early on in the declining phase, which reduces the occurrence even of weak HSS/CIR storms.

On the other hand, during strong solar cycles, there is more of new magnetic flux emerging on the solar surface in the form of sunspots and other active regions.
Accordingly, there are more surges that strengthen the polar fields and form coronal hole extensions, which can occur even in the late declining phase of the cycle.
These changes lead to a later maximum of HSS/CIR storms in these cycles. 
Strong polar fields also push the HCS region thinner, which allows the occurrence of weak HSS/CIR storms even around sunspot minimum.
This evolution culminated during the extreme cycle of SC22, when the solar polar fields and the HMF attained their maximum value during the space age \cite{Smith_Balogh_2008, Zhou_2009}.
Note that the field intensity isolines reach $\pm5 \mu$T in 1990s (left plots of Fig. \ref{fig:WSO_HCS_8plots}) but only $\pm2 \mu$T in 2010s (right plots of Fig. \ref{fig:WSO_HCS_8plots}).
The exceptionally strong polar fields in SC22 produced an exceptionally narrow HCS during the subsequent declining phase and minimum \cite{Richardson_1997}, which allowed the HSS streams to occasionally reach the Earth even during the sunspot minimum.
Quite appropriately, the only cycle when the maximum of weak HSS/CIR storms was on a minimum year, was cycle 22.
After SC22, the weakening sunspot activity produced weaker solar polar fields \cite{Smith_Balogh_2008, Zhou_2009} and smaller polar coronal holes \cite{Harvey_2002, Kirk_2009} in SC23 and SC24, which led to a thicker HCS during the respective declining phases and minima \cite{Virtanen_Mursula_2010}.
This is also seen in the fact that, during this millennium, the Earth has spent an increasing fraction, roughly half of the time within the HCS/streamer belt region.
Also, the slow wind of the streamer belt has become a more important source of weak magnetic storms than ever during the space age.

\section{Discussion and conclusions}
\label{Sec: Sec11 Conclusions}

% MAHDOLLISESTI KÄYTETTÄVIÄ TEKSTEJÄ 

% Coronal holes at solar maximum times are small and may be located at a wide range of latitudes, often near active regions [Cranmer, 2009]. 
% The area of low-latitude coronal holes usually peaks around the solar maximum, while the area of polar coronal holes minimizes [e.g., Fujiki et al., 2016]. 
% Small coronal holes are also a source of slow solar wind during solar maxima [Nolte et al., 1976; Zhang et al., 2003].

% The latitudinal distribution of the source regions of CMEs is very broad near solar maxima, but concentrated closer around the magnetic equator near minima [e.g., Hundhausen, 1993]. 
% Near maxima, CMEs originate from both active regions at low latitudes and filament regions (cooler plasma suspended above the photosphere by magnetic fields) at higher latitudes [Gopalswamy et al., 2003]. 

% Gopalswamy et al. [2003] also studied the solar cycle variation of the occurrence rate and mean/median speeds of CMEs observed with LASCO during solar cycle 23. 
% They found that the CME rate varied from about 0.5 per day during the solar minimum to up to 6 per day during maximum. 
% The CME mean speed was seen to increase by a factor of 2 (from 250 km/s to 500 km/s) from minimum to maximum.

We have studied in this paper the occurrence of magnetic storms of different intensities during the whole space age from 1957 until 2021.
We have used both the original storm index, the Dst index, and its recalculated and corrected version, the Dxt index.
These two indices have differences, e.g., in overall normalization and quiet-time levels.
Therefore the yearly storm numbers in the four intensity categories extracted from these two indices slightly differ between each other.
However, even despite a small systematic long-term trend in the ratio of the two indices, there is no significant difference in the relative occurrence of storms of different intensities according to the two indices.
Rather, we find that all of the main results on the long-term occurrence of magnetic storms and their implications about the evolution of solar magnetic fields and the solar wind are the same using either of the two indices.
Moreover, the changes in the Sun during the space age leading to a varying number of magnetic storms are much larger than the differences in the storm numbers between the two indices.

There were in total 2526 magnetic storms during the space age according to the Dxt index and 2743 storms, i.e., some 8.6\% more, according to the Dst index. 
This implies that there were, on an average, 39/42 storms per year, i.e., roughly three storms per solar rotation.
About 45\% of all storms were weak storms (-50 nT$<$Dxt/Dst$\leq$-30 nT), 40\% moderate storms (-100 nT$<$Dxt/Dst$\leq$-50 nT), 12\% intense storms (-200 nT$<$Dxt/Dst$\leq$-100 nT) and 3\% major storms (Dxt/Dst$\leq$-200 nT). 
So, roughly speaking, almost a half of storms were weak storms and three quarters of the rest were moderate storms.
Almost exactly the same percentages are found for both indices.
The two indices gave also very closely similar storm peak mean values of about -38\,nT, -68\,nT, -131\,nT and -277/-276\,nT for weak, moderate, intense and major storms. 						

We also used solar wind flow type information \cite{Richardson_JGR_2000,Richardson_GA_2012} in order to assign magnetic storms occurring in 1964-2021 to their three main solar wind drivers, the coronal mass ejections (CME), the high-speed solar wind streams (HSS/CIR) and the slow wind region related to the streamer belt and the heliospheric current sheet (HCS).
The HSS/CIR streams produced a bit more than one thousand (1012/1129) solar wind-classified Dxt/Dst storms, almost exactly a half of all solar wind-classified storms in 1964--2021.
There were 785/800 CME storms, making a good third (38.3\%$/$35.7\%) of all solar wind-classified storms, and roughly one CME storm per solar rotation, on an average.
Slow solar wind produced some 300 (262/315) storms, making 12.7\%$/$14.0\% of all solar wind-classified storms, roughly one slow wind storm in a couple of rotations.

The three solar wind streams contributed very differently to the different storm intensity categories. 
CME streams were the cause of all the 48/51 major storms (Dxt/Dst$\leq$ -200 nT) occurring in 1964-2021.
Although CME storms may also include an effect of a HSS, this result proves that the generation of a major storm without a CME was extremely unlikely during the space age.
CMEs produced 84.3\%/84.8\% of intense (-200 nT$<$Dxt/Dst$\leq$-100 nT) storms, HSS/CIR streams produced 14.5\%/14.4\% of them and slow wind streams only 1.2\%/0.8\%. 
Moderate storms were caused slightly more often by HSS/CIR streams (47.5\%/49.9\%) than by CME (45.6\%/41.7\%) streams, or by slow wind (6.9\%/8.4\%).
HSS/CIR streams (61.8\%/62.1\%) were clearly the dominant source of weak storms, but even the slow wind streams (21.3\%/23.0\%) produced more weak storms than CMEs (16.9\%/14.9\%).

The whole solar-terrestrial environment during the space age is characterized by a slow, but unsteadily evolving decrease of solar magnetic activity after the maximum of solar cycle 19, the peak of the Modern Grand Maximum (MGM).
This long-term evolution also affects the sources, occurrence frequencies, intensities and other properties of magnetic storms.
Note also that, since the three main solar wind drivers of magnetic storms depend on different solar parameters and vary at different time scales, the long-term decrease in solar activity affects differently on storms of different origin or different intensity.

Coronal mass ejections are mainly produced by fairly new magnetic flux emerging on the solar surface.
This emergence is evidenced and traditionally even quantified by sunspots.
It has been shown that the occurrence of CMEs follows the sunspot cycle \cite<see, e.g.,>[]{Webb_and_Howard_1994, Gopalswamy_2004, Cremades_2007, Robbrecht_ASR_2006, Webb_LRSP_2012}.
Here we verified that CMEs producing moderate and, separately, large (intense or major) magnetic storms vary closely with the changing sunspot activity, not only over the solar cycle but even at longer time scales.
Large storms, which are almost exclusively produced by CMEs, occurred most frequently in the four years (1957--1960) of the maximum and early declining phase of SC19, the peak of the MGM.
CME storms mainly follow sunspots within a year, thus being produced by fairly newly emerged flux, rather than distributed flux.
% Studying the relation between sunspots and CME storms in more detail, we found that there is an excellent relation between sunspots and the yearly number of all CME storms, and also  separately for large, moderate and weak CME storms.

Sunspots explain typically 60\% of the variation of the yearly number of large and moderate CME storms.
On an average, the yearly number of large CME storms is zero if no sunspots exist, suggesting that active regions without sunspots are not effective in producing large CME storms.
The yearly number of large CME storms increases to 4--5 storms for an intermediate sunspot number of 100 (version 2.0, \cite{SIDC_2022}).
On the other hand, about two moderate CME storms in a year can occur even if no sunspots exist.
This suggests that, e.g., active regions without sunspots, filaments and other forms of distributed (not newly-emerged) solar magnetic fields can produce CMEs that are sufficiently strong for moderate storms.
Increasing yearly sunspot number to 100 (200) increases the yearly number of moderate CME storms, on an average, to about 7 (13) storms.

We studied the correlation between sunspots and yearly number of CME storms separately for each of the five full solar cycles (SC20--SC24) included within the space age.
Most cycles had significant correlations between sunspots and the yearly number of large and, separately, of moderate CME storms.
For large CME storms in SC21 and SC24 correlation was only marginal, which 
%We noted that the less strong correlation during SC21 and SC24 
is due to the fact that large CME storms have two separate peaks around the respective sunspot maxima, with a deep minimum in between. 
This evolution reminds of the so-called Gnevyshev gap which is quite a common feature in the solar cycle evolution of several solar and heliospheric parameters \cite{Gnevyshev_1977, Storini_2003}.
Since sunspots typically depict only a small decrease during a possible Gnevyshev gap (and hardly any in SC21), this leads to a couple of years where storm numbers are small but sunspot numbers high, which deteriorates the overall correlation.
Because such a two-peak structure does not seem to be common to all cycles, it likely results from random fluctuations of rather small numbers of yearly large CME storms.
This will be studied later in more detail.
For moderate CME storms the storm numbers are larger, which reduces the effect of random fluctuations.
Moreover, as discussed above, moderate CME storms can be produced even without sunspots. 
These facts may explain the larger number of moderate storms even at the corresponding Gnevyshev gap minimum, which makes the correlation between sunspots and moderate CME storms more significant even for SC21 and SC24, although a (less deep) Gnevyshev gap is seen also in the number of moderate storms for both cycles. 

The best correlation between sunspots and large and, separately, moderate CME storms was found during cycle 22.
During SC22, sunspots explain 94\% and 77\%, i.e., far more than the average 60\%, of the variation of yearly number of large and moderate CME storms, respectively.
Cycle 22 also marks the highest peak in the number of large CME storms during the solar wind flow covered period (SC20--SC24).
Sunspots depict a clear (but not very deep) Gnevyshev gap during this cycle, and large CME storms follow this evolution closely, without an excessive decrease between the two peaks.
Thus, the cycle evolution of sunspots and large CME storms is very similar during SC22, leading to the extremely high correlation.
On the other hand, the Gnevyshev gap of moderate CME storms in SC22 is relatively deeper than in large CME storms.
This reduces the respective correlation with sunspots in SC22 below that for large storms, but is still the best for moderate storms among all cycles. 
% Accordingly, correlation for moderate CME storms in SC22 remains below that for large storms, but is still the best for moderate storms among all cycles. 

The parameters of the linear correlation between sunspots and large and moderate CME storms also vary considerably from cycle to cycle.
For large CME storms, SC20 and SC23 have almost the same slope as the overall best fit line for all years. 
SC21 and SC24 have considerably lower slopes, while SC22 has a slope more than twice higher than SC21 and SC24.
For moderate CME storms, the five slopes deviate less from each other and from the common mean, probably because of better statistics due to a larger number of storms.
However, despite some difference in correlation parameters, the best fit lines of all cycles remained within the 95\% CL boundaries of the overall fit.
%Nevertheless, although the correlation parameters varied considerably from cycle to cycle, all cycles remained within the 95\%CL boundaries of the overall fit.
Thus, interestingly, no cycle clearly deviated from the other cycles in the relation between sunspots and large or moderate CME storms.
This applies also to the low cycle 24.

However, when studying the relation between sunspots and all CME storms around sunspot minima (using 3-year means in order to increase statistics), we found that the minimum between cycles 23 and 24 deviates from the other four minima by remaining below the 95\% CL correlation boundary.
This minimum was very weak in sunspot activity, although still slightly more active than the following minimum.
However, it was exceptionally weak in CME storms, with no CME storms at all in two years 2007--2008. 
% This gives additional evidence for the view that the CME activity in the minimum after cycle 23 is exceptionally weak, but also shows that it is .
% Our results not only provide additional evidence for the exceptionally weak CME activity around this minimum, but underline its uniqueness among all other solar minima during the space age in breaking the correlation between sunspot activity and CME storm occurrence.
Although this minimum breaks the common correlation between sunspot activity and CME storm occurrence at sunspot minima, the remaining years of the two annexing cycles, SC23 and SC24, recover this relation for full cycles and make it agree with all other cycles.

We have shown in this paper that there is an intimate connection between the occurrence of HSS/CIR storms and the structure of the heliospheric current sheet, in particular its latitudinal width (thickness) and the tilt (curviness).
A thin HCS makes large gradients of solar wind properties around the neutral line (center of the HCS region), whence the Earth is more often affected by high-speed solar wind streams.
On the other hand, the tilt of the HCS determines how high (northern or southern) heliomagnetic latitudes the Earth can reach.
A large tilt in the HCS can produce an intrusion of high-speed solar wind to low heliographic latitudes and to the ecliptic.
The tilt angle of the HCS decreases fairly systematically from very high (about $90^\circ$) to very low tilt angles during the declining phase of the solar cycle.

% FIGURE11
\begin{figure}[t!]
	\centering
	\includegraphics[width=1\linewidth]{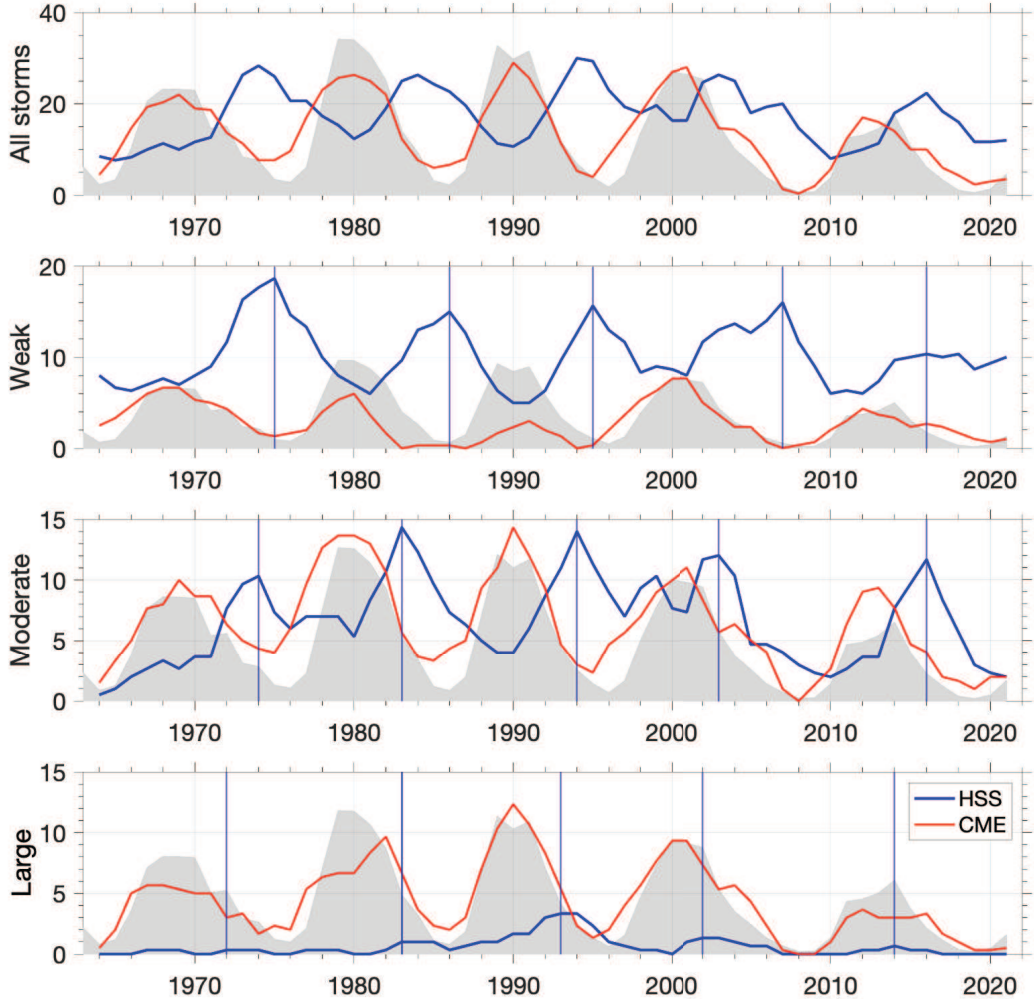}
	\caption{Three-year running mean numbers of CME (red line) and HSS/CIR (blue line) Dxt storms. 
	Panels depict (from top to bottom) all storms, weak storms, moderate storms and large storms.
	Vertical lines denote the locations of cycle maxima of HSS/CIR storms of respective intensity. 
	Annual sunspots are depicted as gray background in each panel. 
	}
	\label{fig:yearly_Dxt_storms_3yr_smooth_KM}
\end{figure}

This decrease of the HCS tilt angle controls the occurrence of HSS/CIR storms of different intensity in the declining phase of all solar cycles.
Due to this decrease, intense HSS/CIR storms tend to occur not later than moderate HSS/CIR which, again, tend to occur not later than weak HSS/CIR storms.
We could see this ordering already in the yearly number of HSS/CIR storms depicted in Figures \ref{fig:yearly_Dxt_storm_sw_numbers_all_3intens} and \ref{fig:yearly_Dst_storm_sw_numbers_all_3intens}.
However, since random fluctuations may have some effect on the ordering of peaks and, thereby, cast doubt on our results, we have calculated three-year running means of the yearly CME and HSS/CIR storm numbers, separately for weak, moderate, large and all Dxt storms, and depicted them in Figure \ref{fig:yearly_Dxt_storms_3yr_smooth_KM}.
Three-year running means weight the distribution of storms more widely and, therefore, are less vulnerable to  random fluctuations of yearly peaks.
We have included in Figure \ref{fig:yearly_Dxt_storms_3yr_smooth_KM} also vertical lines to indicate the cycle maxima of HSS/CIR storms of each intensity class.
(In case two years had the same mean number, the first was marked; this only applies to large HSS/CIR storms where numbers are small). 

The vertical lines of Figure \ref{fig:yearly_Dxt_storms_3yr_smooth_KM} clearly demonstrate the above discussed ordering of the cycle maxima of HSS/CIR storms.
During all the five cycles SC20--SC24 of space age, the cycle peak of large HSS/CIR storms did not occur later than the corresponding peak of moderate HSS/CIR storms which, again, was not later than the peak of weak HSS/CIR storms in the same cycle.
This ordering is most clear in the declining phase of SC22, where the three-year mean maxima of large, moderate and weak HSS/CIR storms follow each other in steps of one year from 1993 to 1994 and 1995.
As noted earlier, cycle 22 is also exceptional in the number of large HSS/CIR storms, which is dramatically clearly seen in the bottom panel of Figure \ref{fig:yearly_Dxt_storms_3yr_smooth_KM}.
More than 40\% of all large HSS/CIR storms in 1964--2021 occurred in the six years of 1991--1996.
Although neither moderate nor weak HSS/CIR storms attained their space age peak during SC22, the three-year running mean number of all HSS/CIR storms maximized in 1994, in the late declining phase of SC22.
These results are due to the exceptionally thin HCS in the declining phase of SC22 \cite{Richardson_1997}, which resulted from exceptionally strong solar polar and heliospheric fields in the declining phase of this cycle \cite{Smith_Balogh_2008, Zhou_2009}.
Note also that the solar magnetic polarity is positive around the minimum between cycles 22 and 23, which enhances geomagnetic activity at high heliolatitudes (in Spring and Fall) by the Russell-McPherron (RMP) mechanism \cite{Russell_McPherron_1973}.
While this mechanism alone does not cause the exceptional storminess of cycle 22, the thin current sheet also increases the occurrence of dominant HMF sectors at high heliolatitudes, (the so-called Rosenberg-Coleman effect \cite{Rosenberg_Coleman_1969}), which enhances the effectivity of the RMP mechanism during cycle 22.
Thus, the increased HSS activity and the enhanced RMP effectivity, both due to the exceptionally thin HCS, lead to the exceptional occurrence of large HSS/CIR storms during the declining phase of cycle 22.

Similarly as for CME storms, the long-term decrease of sunspot activity during the space age also affects the occurrence of HSS/CIR storms.
However, the decrease during the space age was not steady or continuous, neither in sunspot activity nor in magnetic storms.
The total number of storms experienced two dramatic dropouts, one in the declining phase of SC19, from the record storm level in 1957--1960 until the second-lowest minimum in 1965, and the other in the declining phase of SC23, which continued to the low cycle 24 (see Fig. \ref{fig:yearly_storm_numbers_4intens}).
Between these two dropouts, both sunspot and storm activity first increased from SC20 on until cycle 22, and then slightly decreased in SC23.
The increase in sunspot activity from SC20 on strengthened the intensity of polar magnetic fields, which culminated in SC22 \cite{Smith_Balogh_2008, Zhou_2009}.
Strong polar fields pressed the HCS region exceptionally thin in the declining phase of SC22, producing a record of large HSS/CIR storms during the space age.

Since SC22, polar fields have continued to decline, which has widened the HCS region.
During a wide HCS, the Earth can reach the fast solar wind only at times when the HCS tilt angle is large.
Therefore, the maxima of HSS/CIR storms occur earlier in the declining phase of weak cycles of SC23 and SC24 than during the earlier, stronger cycles. 
This can be seen in the yearly storm numbers of Figures \ref{fig:yearly_Dxt_storm_sw_numbers_all_3intens} and \ref{fig:yearly_Dst_storm_sw_numbers_all_3intens}, and it can even more clearly be seen in the three-year running mean storm numbers of Figure \ref{fig:yearly_Dxt_storms_3yr_smooth_KM}.
The shift of cycle peaks of HSS/CIR storms from the late declining phase in the more active cycles SC20--SC22 to the early declining phase in the recent, more quiescent cycles SC23 and SC24 is, very appropriately, most clearly visible for the large HSS/CIR storms. 
In SC20--SC22, the maxima of large HSS/CIR storms are found some 3--4 years after the respective sunspot maximum, but in SC23 only 1--2 years later and in SC24 the two maxima occur in the same year.

Since the cycle maxima of moderate and weak HSS/CIR storms occur, as discussed above, later (or not at least earlier) than large HSS/CIR storm peaks, this long-term shift is less clearly seen for the weaker HSS/CIR storms.
For moderate storms, this shift is still seen even in the smoothed numbers of Figure \ref{fig:yearly_Dxt_storms_3yr_smooth_KM} (peaks in SC20--SC22 are 4--6 years, in SC23 2--3 years and in SC24 2 years after the respective sunspot maximum).
Note, first, that the 3-year smoothing moves the weak storm peak in SC23 to 2007, while for yearly storm numbers it was in 2003 (see Figs. \ref{fig:yearly_Dxt_storm_sw_numbers_all_3intens} and \ref{fig:yearly_Dst_storm_sw_numbers_all_3intens}).
Year 2003 was the peak storm year for both weak and moderate (and, naturally, all) HSS/CIR storms, and only one year after the peak in large HSS/CIR storms.
This year is known to be the record year of geomagnetic activity in space age \cite{Mursula_2015, Mursula_GRL_2017} and one of the record years of HSS occurrence (see Fig. \ref{fig:yearlySWTypeFractions}).
In this year, the HSSs mainly originated from low-latitude coronal holes \cite{Gibson_2009, Mursula_GRL_2017, Hamada_SP_2021}.
Accordingly, the occurrence of HSS/CIR storms at this time was less strongly controlled by solar polar coronal holes than in the other cycles.

Secondly, the occurrence of weak HSS/CIR storms in SC24  (and partly even in the declining phase of SC23) was increasingly affected by the widening of the HCS.
As noted earlier, the yearly number of weak HSS/CIR storms in the minimum of SC24, and later in the early ascending phase of SC25, is not much less than in the prior declining phase (see Figs. \ref{fig:yearly_Dxt_storm_sw_numbers_all_3intens} and \ref{fig:yearly_Dst_storm_sw_numbers_all_3intens}).
This leads to the fairly constant (smoothed) number of weak HSS/CIR storms in Figure \ref{fig:yearly_Dxt_storms_3yr_smooth_KM}) since the maximum of SC24.
Thirdly, since the HCS has been exceptionally wide for about 15--20 years, the Earth stays, for the same tilt angle, longer inside the HCS region.
Thus, the decrease of solar activity and solar polar fields since SC22 have led to the fact that the Earth stays, since the declining phase of SC23, more than 50\% of the time within the HCS and the related slow solar wind (streamer belt).
This has greatly reduced the number of intense and moderate HSS/CIR storms, and significantly raised the relative fraction of slow solar wind as the source of weak magnetic storms.

Concluding, we have studied here the occurrence of magnetic storms of different intensity over the space age (1957--2021).
Space age is characterized by a slow, unsteady decrease of solar activity, which directly affects the CME storms and, indirectly, via its effects to the structure of the heliospheric current sheet, also the HSS/CIR storms.
We find that the variations in sunspot activity closely explain the variation in the yearly number of CME storms over the whole space age.
All cycles separately comply to this common, rather tight connection between sunspots and CME storms.
However, during the weak solar minimum between cycles 23 and 24 the number of CME storms is smaller than predicted by the common relation.
During four years from 2006--2009, there were only five CME storms, out of which only one was large.
In comparison, there were 12 CME storms in the second-weakest four-year time interval of 2018--2021.
Increasing sunspot activity from SC20 to SC22 increased solar polar fields, which led to an exceptionally thin HCS during the declining phase of SC22, producing a record of large HSS/CIR storms.
Subsequent decrease in sunspot activity has weakened solar polar fields and widened the HCS region.
These long-term changes in HCS width have affected the occurrence of HSS/CIR storms during the solar cycle so that HSS/CIR storms occur, since SC23, in the early to mid-declining phase of the solar cycle, while during the earlier, more active cycles 20--22, they maximized in the late declining phase.
Widening HCS has also increased the significance of the HCS and the related slow solar wind (streamer belt) for the Earth.
The Earth spends now, since the declining phase of SC23, i.e., roughly from the start of the new millennium, about 50\% of the time, more than ever before during the space age, in the slow solar wind.
This has reduced the number of large and moderate HSS/CIR storms and increased the fraction of slow wind as the source of weak magnetic storms.

\acknowledgments
The authors acknowledge financial support by the Academy of Finland
to the Postdoctoral Researcher project of L. Holappa (No. 322459) and to the PROSPECT project (No. 321440) of T. Asikainen. 
T. Qvick was supported by University of Oulu Graduate School. 
We thank I. Richardson for the updated version of solar wind flow data \cite{Richardson_GA_2012}.
The Dxt index is maintained by the University of Oulu at http://dcx.oulu.fi/ \cite{Karinen_Mursula_2005}.
The Dst index is provided by the WDC for Geomagnetism, Kyoto, Japan at http://wdc.kugi.kyoto-u.ac.jp/wdc/Sec3.html \cite{WDC_C2_Dst_official}.
Sunspot data are from the World Data Center SILSO, Royal Observatory of Belgium, Brussels, on-line catalogue at https://www.sidc.be/silso/datafiles.
The WSO source surface maps are available at http://wso.stanford.edu/synsourcel.html.
% Solar wind data (solar wind speed and different components of HMF) were downloaded from the OMNI2 database (http://omniweb. gsfc.nasa.gov/).

%% ------------------------------------------------------------------------ %%
%% References and Citations

%%%%%%%%%%%%%%%%%%%%%%%%%%%%%%%%%%%%%%%%%%%%%%%
%
% \bibliography{<name of your .bib file>} don't specify the file extension
%
% don't specify bibliographystyle
%%%%%%%%%%%%%%%%%%%%%%%%%%%%%%%%%%%%%%%%%%%%%%%

\bibliography{storm_references}

\end{document}